\documentclass[aip,pof,amsmath,amssymb,preprint]{revtex4-1}
\usepackage{graphicx}
\usepackage{dcolumn}
\usepackage{bm,soul}
\usepackage{siunitx}
\setstcolor{red}
\usepackage[utf8]{inputenc}
\newcommand*{\threeemdashes}{\textemdash\kern-1pt\textemdash\kern-1pt\textemdash}
\newcommand*{\D}{\text d}
\usepackage{xcolor}
\definecolor{dark-green}{rgb}{0,0.5,0}
\definecolor{dark-red}{rgb}{0.8,0,0}
\definecolor{dark-blue}{rgb}{0,0,0.9}
\usepackage{amsmath}

\begin{document}

\preprint{Ansari-Bhateja-Sharma-2022}

\title{Axial segregation of  granular mixtures in laterally shaken multi-trapezium channels}

\author{Mohammed Istafaul Haque Ansari}
\email{istafaul.haque86@gmail.com}
\affiliation{Department of Mechanical Engineering, Indian Institute of Technology Kanpur  208016, Uttar Pradesh, India}

\author{Ashish Bhateja}%
\email{ashish@iitgoa.ac.in}
\affiliation{School of Mechanical Sciences, Indian Institute of Technology Goa, Ponda 403401, Goa, India}%

\author{Ishan Sharma}%
\email{ishans@iitk.ac.in}
\affiliation{Department of Mechanical Engineering, Indian Institute of Technology Kanpur 208016, Uttar Pradesh, India}%

\date{\today}

\begin{abstract}
We investigate axial segregation of binary mixtures  in a laterally shaken horizontal channel formed by ratchet-like sidewalls that appear as concatenated trapeziums when not offset axially. Grain mixtures shaken in such a channel are observed to segregate in two stages: they first separate rapidly into two vertically arranged layers and, then, these layers move axially in opposite directions,  segregating the two  species. Here, we conduct experiments to study the influence on the segregation process of various parameters: the size ratio of grains, the shaking frequency and the channel's geometry. We find that (a) segregation quality depends upon shaking frequency and it is possible to find a unique optimal frequency for segregation, (b) the optimal frequency lowers with increase in size ratio, (c) segregation is generally poorer when the sidewalls are more inclined to each other, and (d) segregation is improved when the sidewalls are axially offset from each other. We then carry out discrete element simulations of the segregation process in order to relate the experimental observations to the interfacial pressure gradient mechanism of Bhateja et al. [{Bhateja, A., I. Sharma and J. K. Singh 2017. Segregation physics of a macroscale granular ratchet,} Phys. Rev. Fluids {\bf 2}, 052301]. We demonstrate that the segregation quality correlates  well with the scaled interfacial pressure gradient, which is the  ratio of the interfacial pressure gradient between the layers, formed in the first stage of the segregation process, to the axial body force provided by the tapered sidewalls.  
\end{abstract}

\maketitle


\section{Introduction}
\label{sec:intro}
Granular materials are typically composed of constituent grains with different sizes, densities, and surface properties. One of the many intriguing and practically relevant features of such materials is segregation, wherein an originally uniform mixture comprising grains of different geometrical and material properties demixes into its constituent species when externally energised\cite{ottino_khakhar2000,kudrolli2004,gray2018}. Segregation of granular mixtures is observed in nature and industry in a variety of granular systems, e.g. vertically\cite{rosato87,shinbrot1998,liao2014} and horizontally\cite{schnautz2005,mobarakabadi2013,bhateja2017,mobarakabadi2017} vibrated containers, rotating cylinders\cite{ottino_khakhar2000,richard2008,yang2017a,yang2017b}, and flow on an inclined surface\cite{gray2018,staron2014,mandal2019}. A comprehensive understanding of granular segregation remains elusive, given its dependence on a number of factors such as size\cite{kudrolli2004}, density\cite{tripathi2013} and shape\cite{mandal2019} differences of grains, interstitial fluid\cite{yan2003}, and shape of the container\cite{grossman1997}. Several mechanisms\cite{rosato87,savage88,knight1993,fan2011,bhateja2017} and theories\cite{gray2018,umbanhowar2019} for segregation have been reported, including percolation\cite{rosato87} where voids between big grains make way for small ones to settle down, kinetic-stress based mechanism\cite{fan2011} causing the motion of small grains from a high to a low  fluctuational velocity region, development of convection currents\cite{knight1993}, and an interfacial pressure-gradient mechanism\cite{bhateja2017}.

\begin{figure}[t]
\centering
\includegraphics[scale=0.25]{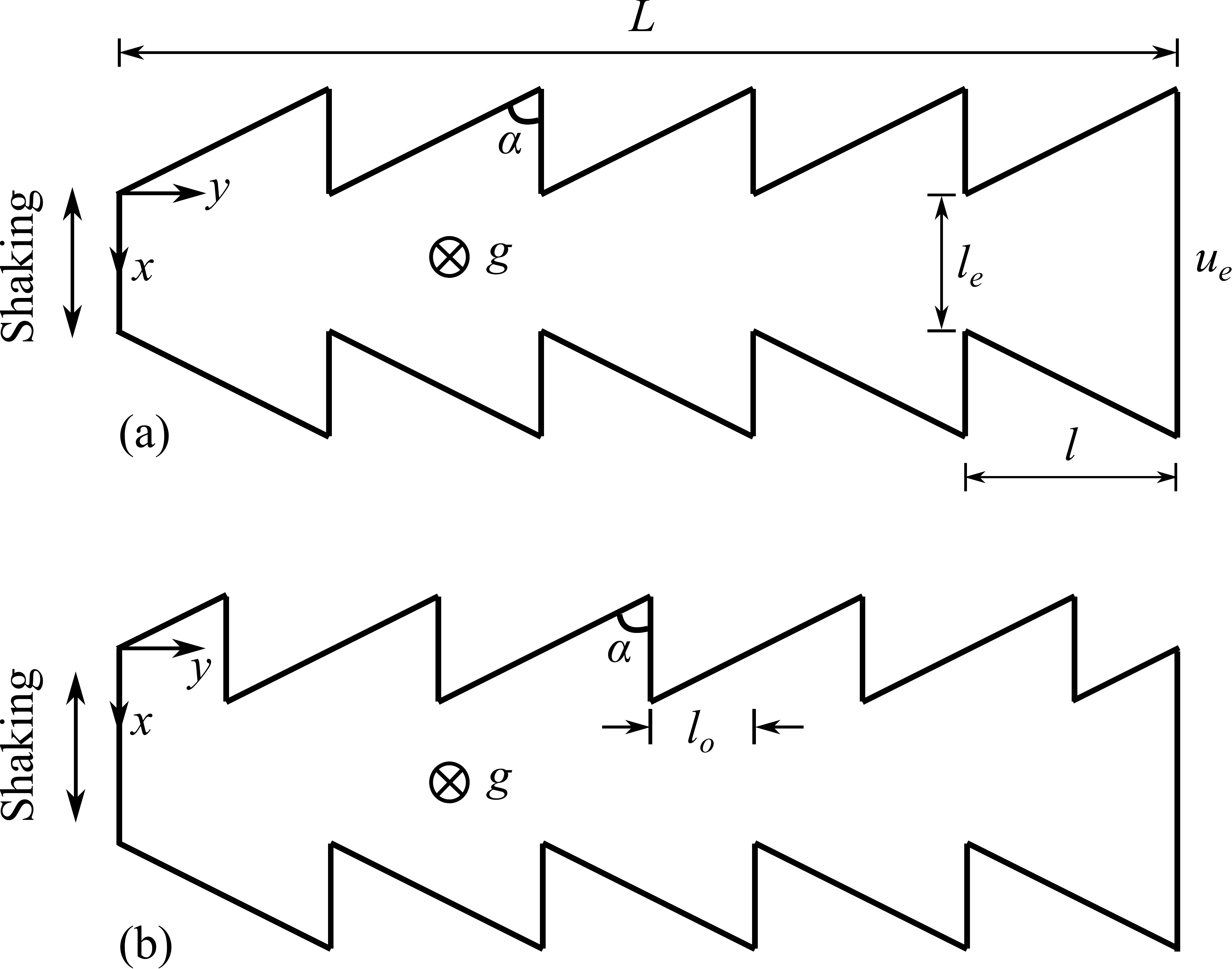}
\caption{Schematics showing the top views of the (a) regular and (b) staggered multi-trapezium channels. The acceleration due to gravity $g$ acts normal and into the channel base. Coordinate axes $x$ and $y$ are aligned with the shaking direction and channel axis, respectively. The taper angle $\alpha$ and the offset $l_o$ are also shown, and $l$ represents the length of a trapezoidal section for regular channel.}
\label{fig:setup}
\end{figure}
\begin{figure}[t]
\centering
\includegraphics[scale=0.3]{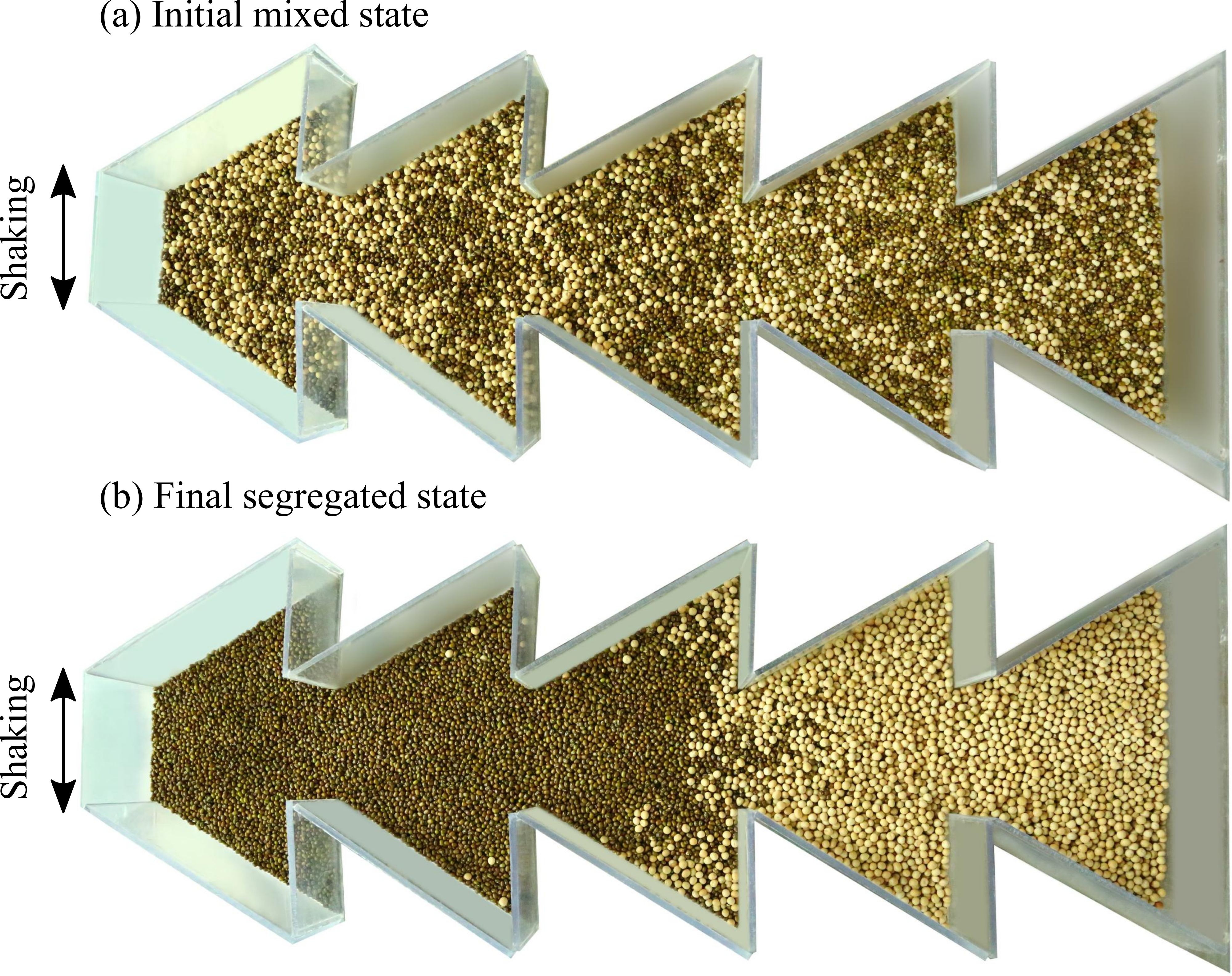}
\caption{An example of the segregation achieved in a laterally shaken multi-trapezium channel. This figure is taken from Bhateja et al. \cite{bhateja2017}.}
\label{fig:expt}
\end{figure}

The {\em interfacial pressure-gradient mechanism} was proposed recently by Bhateja \textit{et al.}\citep{bhateja2017} to explain the axial segregation in  binary mixtures of grains when they are shaken laterally in a horizontal, multi-trapezium, closed-end channels, such as those shown schematically in Fig.~\ref{fig:setup}. They reported that the segregation in such systems proceeds by rapid vertical sorting of the constituent species normal to the channel's base, followed by slow  separation of the species in the axial ($y$) direction. Figure~\ref{fig:expt} shows an initial  mixture and the final outcome of a typical experiment. 

\begin{figure}[t!]
\centering
\includegraphics[scale=0.45]{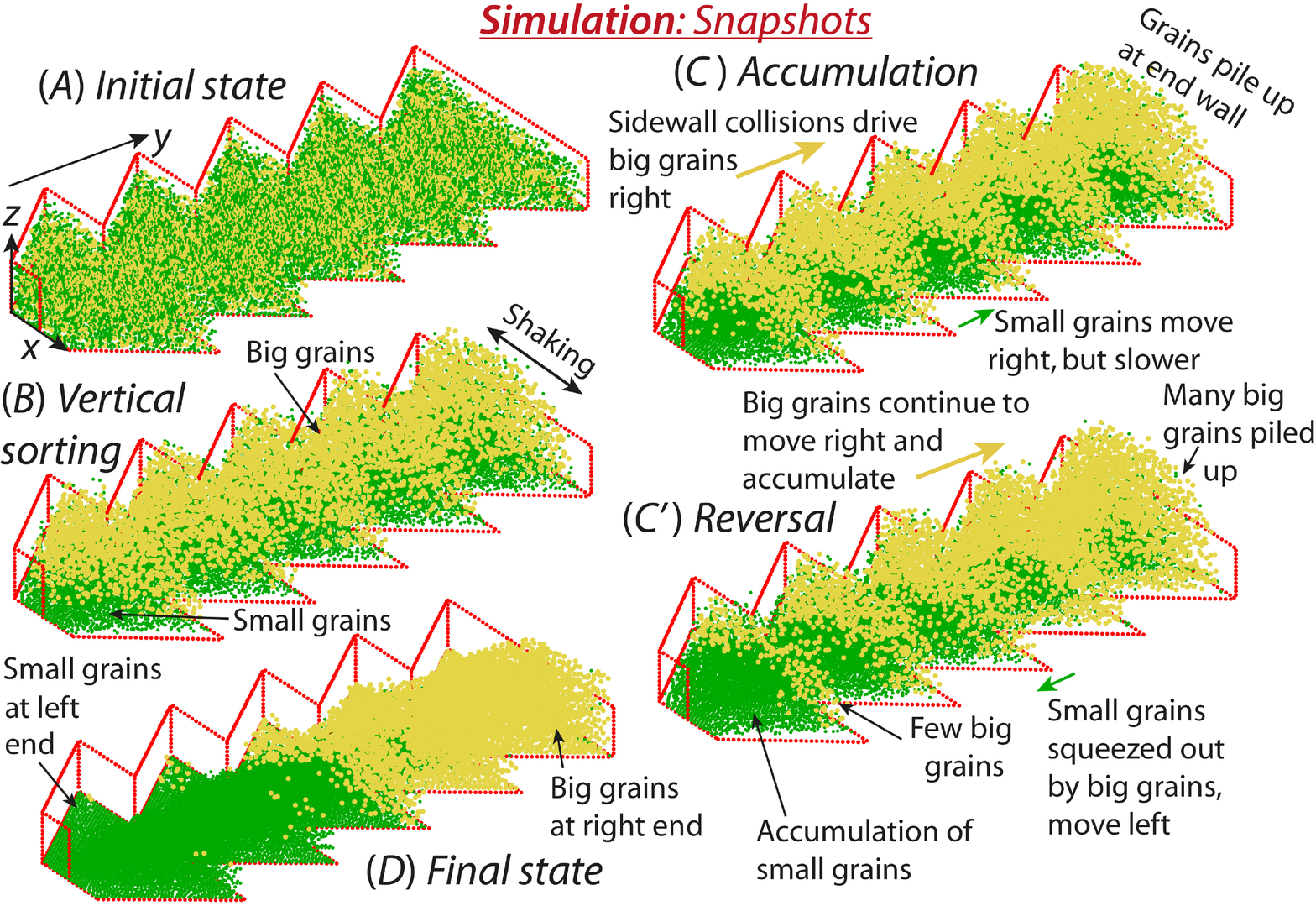}
\caption{Three-dimensional visualization of a discrete element simulation of Bhateja et al. \cite{bhateja2017} that we utilize to explain axial segregation in a laterally shaken multi-trapezium channel. See text for discussion. 
This figure is taken from Bhateja et al.\cite{bhateja2017}.}
\label{fig:cartoon}
\end{figure}


We now appeal to the three-dimensional plots in Fig.~\ref{fig:cartoon} obtained from the discrete element simulations conducted by Bhateja et al. \citep{bhateja2017} to summarize their explanation of the process. When the binary mixture Fig.~\ref{fig:cartoon}(A) is shaken laterally in a multi-trapezium channel, vertical sorting -- which, depending upon the relative sizes and densities of the two species, is the outcome of the Brazil-nut \cite{rosato87, mobius2001}  or the reverse Brazil-nut effect \cite{hong2001, shinbrot2004} --  quickly stratifies the grains into two layers; cf. Fig.~\ref{fig:cartoon}(B). The subsequent slow axial separation of the two layers relies upon a gentle {\em interfacial pressure gradient} that develops between the top and bottom layers of grains, as we now describe. Repeated collisions with the slanted sidewalls push both layers towards the right end of the channel, as shown in  Fig.~\ref{fig:cartoon}(C), where they begin to pile up at the end wall. The difference in the mobility of the top and bottom layers causes the grains in the top layer to accumulate faster at the right-end wall. Because of this, there is an increase in the {\em interfacial pressure} between the grains in the top and bottom layers along the channel's axis toward its right end, and this may be observed in Fig.~\ref{fig:cartoon}(C$'$). This  interfacial pressure gradient then squeezes the grains in the bottom layer towards the channel’s left end leading to the final segregated state displayed in Fig.~\ref{fig:cartoon}(D).


This paper aims to extend the study of Bhateja \textit{et al.}\citep{bhateja2017} with the following two objectives. First, we will employ experiments to identify parameters which affect and optimize segregation in the multi-trapezium channel. The parameters investigated here include the shaking frequency, the size ratio of the grains and the channel's geometry.  Second, utilizing soft-sphere discrete element (DE) simulations we will compute the interfacial pressure gradient driving the axial segregation, and correlate it with experiments, in order to provide further confirmation and insight into the segregation mechanism proposed by Bhateja \textit{et al.}

The paper is organized as follows. Details of experiments along with system parameters and procedure for data analysis are given in Sec.~\ref{sec:exp}. We discuss results obtained in experiments and simulations in Secs.~\ref{sec:results-E} and \ref{sec:results-S}, respectively. We conclude  in Sec.~\ref{sec:conclude}.

\section{Experiments}
\label{sec:exp}
\subsection{System and parameters}
\label{sec:exp-a}
Our experimental setup consists of a horizontal channel built of several interconnected trapezoidal sections, as shown in Fig.~\ref{fig:setup}(a). The dimensions of the setup are as follows: $L\approx \SI{90}{cm}$, $l_{e}=\SI{10}{cm}$, and $u_{e}=\SI{30}{cm}$. We also consider a variant of the multi-trapezium channel displayed in Fig.~\ref{fig:setup}(a), in which the trapezoidal sections are offset by an amount $l_o$; see Fig.~\ref{fig:setup}(b). Henceforth, we will call  the former and latter channels as, respectively, the \textit{regular} and \textit{staggered} multi-trapezium channels. The channels are closed at both ends, have a flat base and are open at the top. The slant of the channel sidewalls is given by the {\em taper angle} $\alpha$, as illustrated in Fig.~\ref{fig:setup}.

\begin{figure}
\centering
\includegraphics[scale=0.45]{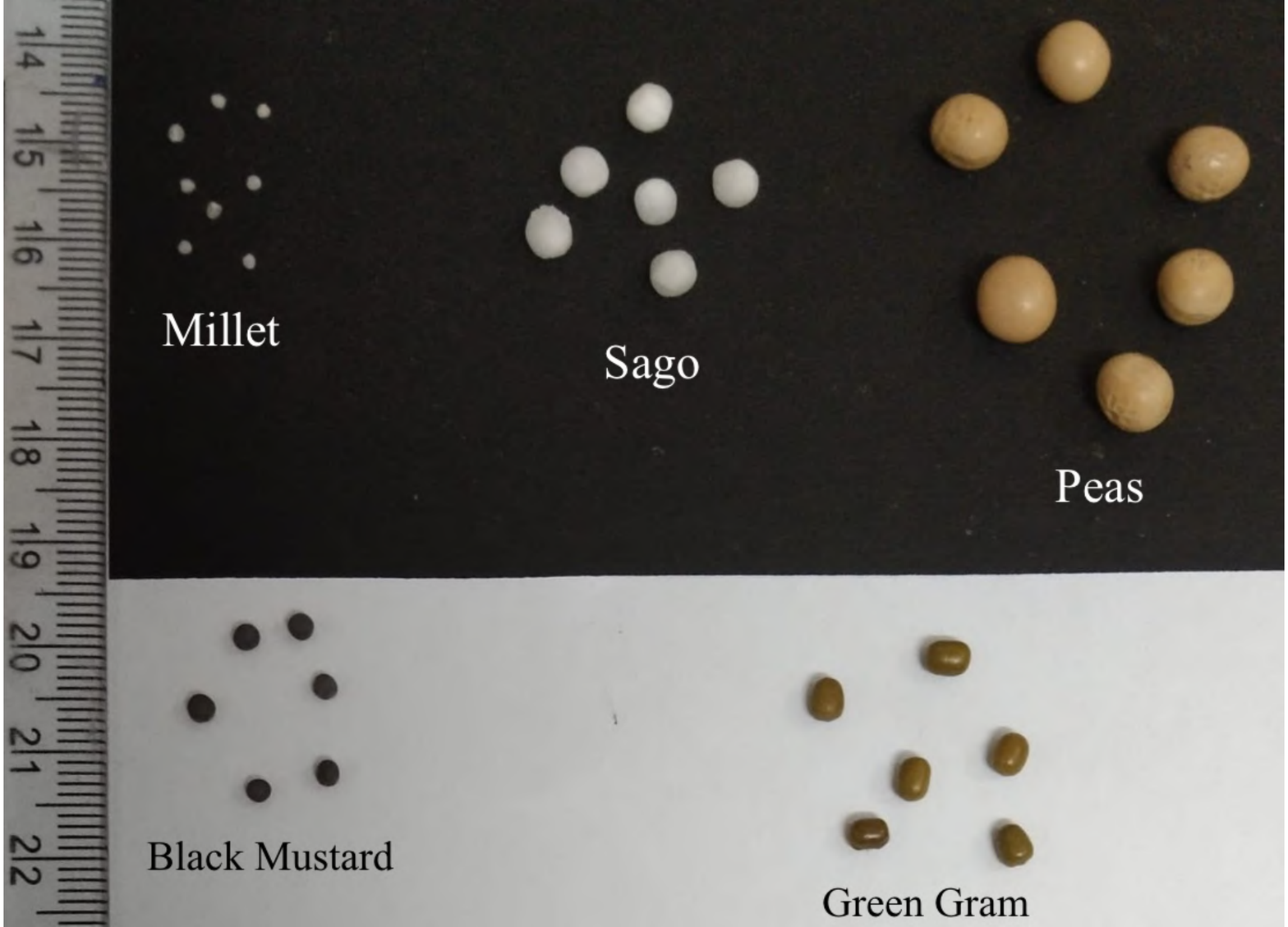}
\caption{A snapshot showing the shape of the foodgrains.}
\label{fig:grains}
\end{figure}

The channels are shaken in a sinusoidal manner in the horizontal plane along the lateral direction, i.e. transverse to the channel's axis, as indicated in  Fig.~\ref{fig:setup}. The experiments are performed at a fixed shaking amplitude $A=7\ \text{cm}$ and at frequencies varying between $N=80\ \text{\text{rpm}}\ (1.33\ \text{Hz})$ and $130\ \text{\text{rpm}}\ (2.167\ \text{Hz})$ in increments of 5 \text{rpm}. Typically, dimensionless acceleration $\Gamma=A (2\pi N/60)^2/g$ is used to quantify the shaking strength of a vibrated granular media. Here, we keep the amplitude fixed and vary the frequency, so that $\Gamma$ varies between 0.5 and 1.32 corresponding to the lowest and highest $N$. For simplicity, the data in this work are presented in terms of $N$, as the amplitude is kept constant. We employ various foodgrains of different sizes and shapes as the granular material in our experiments. The shape, size with standard deviation, and density of the foodgrains are given in Table~\ref{tab:table1}. The shape of the foodgrains are  displayed in Fig.~\ref{fig:grains}.
\begin{table}
\begin{tabular}{|p{2.5cm}|p{2.6cm}|p{3.5cm}|p{3cm}|} 
 \hline
 \textbf{Foodgrains} & \textbf{Shape} & \textbf{Size} (\textbf{SD}) & \textbf{Density} ($\boldsymbol{\rho}$)\\
 \hline
 Peas & Near-spherical & $d \approx \SI{7}{mm}$ (0.413) & \SI{769} {kg.m^{-3}}\\
 Green Gram & Elongated & $d_{min} \hspace{-.25cm}\approx$ \hspace{-.2cm}\SI{3.5}{mm} \hspace{-.2cm}(0.295) $d_{max} \approx$ \SI{5}{mm} (0.515) & \SI{790} {kg.m^{-3}}\\
 Sago & Near-spherical & $d \approx$ \SI{4.1} {mm} (0.347) & \SI{705.6}{kg.m^{-3}}\\
 Black Mustard & Near-spherical & $d \approx$ \SI{2.2} {mm} (0.193) & \SI{720} {kg.m^{-3}}\\
 Millet & Near-spherical & $d \approx$ \SI{1.5}{mm} (0.215) & \SI{640}{kg.m^{-3}}\\
 \hline
\end{tabular}
\caption{Shape, size [with standard deviation (SD)], and density of the foodgrains \citep{mpd,etb}.}
\label{tab:table1}
\end{table}
For analyzing the role of size ratio, we consider four different binary mixtures comprising equal weights of two different types of foodgrains. The size ratio $s_r=d_b/d_s$ and density ratio $\rho_r=\rho_b/\rho_s$ of the mixtures that we investigate are provided in Table~\ref{tab:table2}, where $d$ is the diameter and $\rho$ is the density of grains, and the subscripts $b$ and $s$ correspond, respectively, to big and small grains. We note that the density of the foodgrains is nearly the same. The taper angle  $\alpha$ varies between $15^{\circ}$ and $75^{\circ}$ in steps of $15^{\circ}$.
\begin{table}
\begin{tabular}{|p{4.5cm}|p{3cm}|p{4cm}|} 
 \hline
 \textbf{Mixture type} & \textbf{Size} \textbf{ratio} (\bm{$s_r$}) & \textbf{Density} \textbf{ratio} (\bm{$\rho_r$})\\
 \hline
 Peas and green gram & 1.4 &  0.97 \\
 Millet and black mustard & 1.47 & 1.125\\
 Sago and black mustard & 1.86 & 0.98\\
 Peas and black mustard & 3.18 & 1.07\\
 \hline
\end{tabular}
\caption{Size and density ratios of the binary granular mixtures of foodgrains employed in our experiments.}
\label{tab:table2}
\end{table}
\subsection{Procedure and Data Analysis}
The procedure followed for conducting experiments is as follows. At the beginning of every experiment, a binary mixture is prepared by homogeneously mixing by hand equal weights ($1.2$ kg each) of two  different foodgrains. The experiments are performed until a final segregated steady-state is achieved. The time to achieve the steady-state depends upon the shaking frequency and varies between $20$ to $25$ minutes. The experiment is stopped once the steady-state is achieved, and then the images are captured. We observed that the grains moved synchronously with the channel like a solid body and did not segregate when the system was shaken below $80\ \text{\text{rpm}}$. This reflects a lack of adequate fluidization at low frequencies. Ansari \textit{et al.} \cite{ansari2018} also reported a similar frequency threshold in the Brazil-nut system, which was coupled with the frictional force and weight of the grains. They did not observe any fluidization in the system below this threshold frequency. Similarly, at the other extreme, segregation did not occur when the system was shaken at frequencies larger than $130$\ \text{rpm}. This is because, at high shaking frequencies, the rightwards axial momentum provided by the sidewalls to the bottom layer of grains dominates the pressure gradient that drives this layer leftwards, so that all grains displace towards the right end. For this reason we investigate shaking frequencies lying between 80 and 130 \text{rpm}.

We now define two parameters -- \textit{segregation index} ($\sigma$) and \textit{normalized overlap} ($\lambda$)  -- that we will  employ  to quantify the degree of segregation in our system. The segregation index $\sigma^s$ corresponding to either of the two grain species ($s=1,2$) is defined as
\begin{equation}
\label{eq:index}
 \sigma^s=1-\int_0^L\left(f_y^s-f_{I}^s\right)^2 \D y,
\end{equation}
where $f_y^s$ is the total number of pixels in the image of the system's top view corresponding to grain type $s$ in a slice of length $\D y$ that spans the channel's width, while $f_I^s$ is the pixel count in an \textit{ideally segregated} system, which we define below through two examples. We note that we employed \eqref{eq:index} for computing the segregation index for both types of grains, i.e. for $s=1$ and $2$, and found that $\sigma^1\simeq\sigma^2$. Thus, we define the average segregation index $\sigma=(\sigma^1+\sigma^2)/2$ and report this value henceforth.

\begin{figure}
\centering
\includegraphics[scale=0.5]{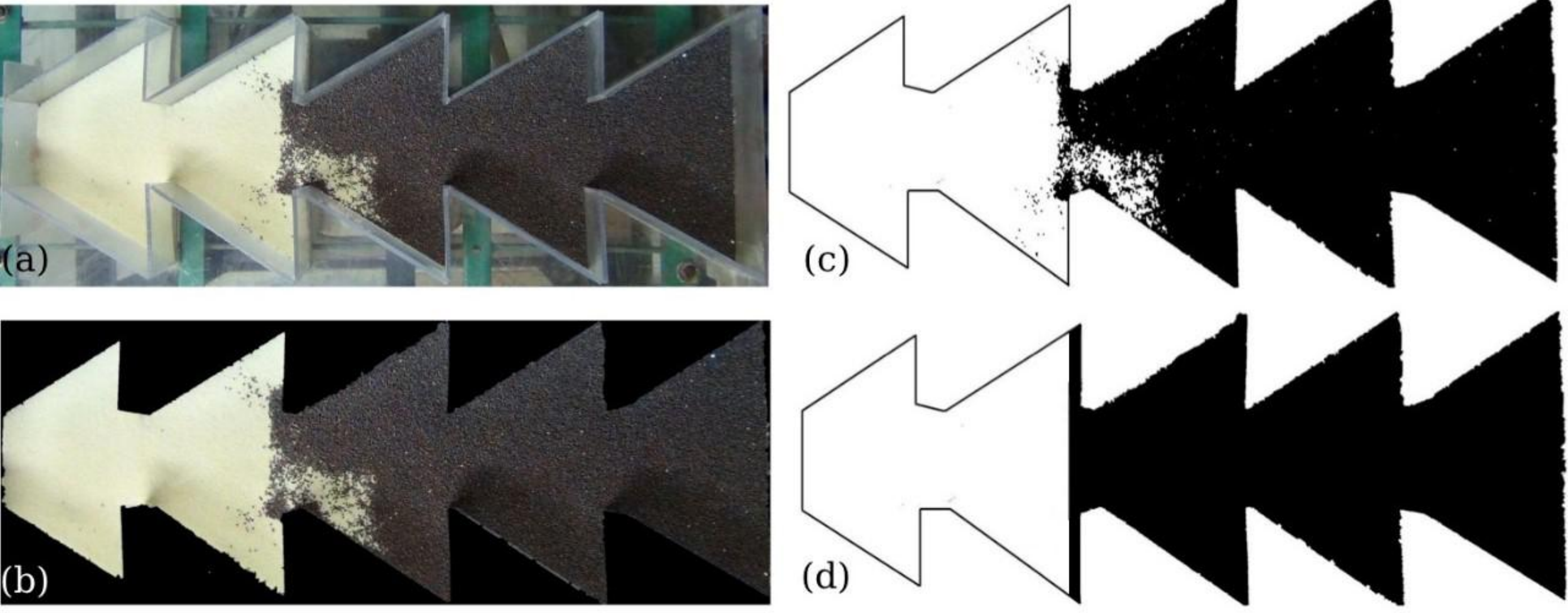}
\caption{Final segregated state of the millet and black mustard mixture at $95$ \text{rpm} in the channel with $\alpha=60^{\circ}$. (a) Raw color image, (b) a cropped version of the raw image after eliminating the sidewalls and background, (c) a binary version of the cropped image showing mustard grains in black, and (d) an ideal binary image generated by identifying the interface in the system and setting every pixel in the image on the right side of the interface to be black and on its left to be white. In (b) - (d) the sharp connection between the first and second trapezium from the left in (a) appears like a neck because of the distortion at the left edges of the top view so that, when the image is cropped, the vertical edge at the sidewall becomes visible and takes the shape of a line. This minor distortion in the image, however, does not affect our calculations.}
\label{fig:image_analysis_1}
\end{figure}

\begin{figure}[b!]
\centering
\includegraphics[scale=0.75]{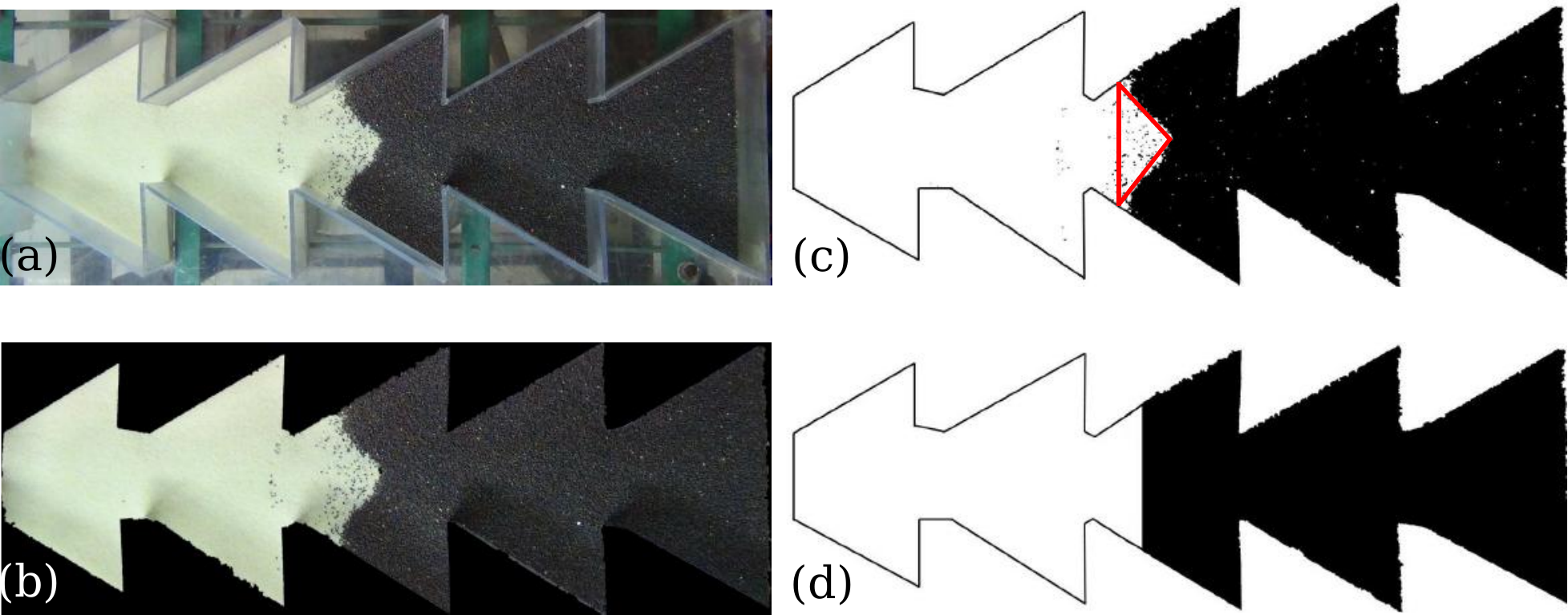}
\caption{Same as Fig.~\ref{fig:image_analysis_1}, but now displaying the final segregated state of the millet and black mustard mixture shaken at $105$ \text{rpm}.}
\label{fig:image_analysis_2}
\end{figure}

Figure~\ref{fig:image_analysis_1}(a) shows an image of a final segregated state of a binary mixture of millet and mustard grains at $95\ \text{\text{rpm}}$ in a channel with $\alpha=60^{\circ}$, where mustard and millet have segregated towards the right and left ends, respectively. The corresponding cropped image is shown in Fig.~\ref{fig:image_analysis_1}(b) after removing the sidewalls and the background. Figure~\ref{fig:image_analysis_1}(c) displays the binary version of Fig.~\ref{fig:image_analysis_1}(b) in black and white, with the colors representing, respectively, mustard and millet. Two trapeziums of the channel are superimposed in the left portion of this binary image to guide the reader's eye. It is evident from Fig.~\ref{fig:image_analysis_1}(c) that there is a discernible linear interface lying along the $x$-axis to the left of which  there are very few mustard grains. At the same time, there is a patch of millet grains to the right of this interface, which appears to be small. We now identify and fix this interfacial line and set every pixel in the image on its right side to be black and on the left to be white to generate a sharp interface, and this transformed image, which corresponds to an \textit{ideally} segregated state, is displayed in Fig.~\ref{fig:image_analysis_1}(d). The function $f_I^s$ in (\ref{eq:index}) corresponds to this ideally segregated version of the experimentally derived image in Fig.~\ref{fig:image_analysis_1}(c), with $s = 1$ or 2 depending upon which species -- millet or black mustard in the present case -- that we focus upon. As another example, consider the segregated state at $105$ \text{rpm} shown in  Fig.~\ref{fig:image_analysis_2}, wherein the interface is observed to be V-shaped. In such scenarios, in order to generate the ideal binary image, we first locate the interface as a lateral line, i.e. lying along the $x$-axis, and passing through the centroid of the triangle constructed in Fig.~\ref{fig:image_analysis_2}(c). We then set every pixel in the image on the right side of this imaginary interfacial line to be black and on the left to be white, as in the preceding example, to obtain the ideally segregated state in Fig.~\ref{fig:image_analysis_2}(d).

We note that more different and/or more complex interfaces could have been chosen to define the ideally segregated state. However, we believe that doing so will defeat our aim for a simple definition for the ideally segregated system. Moreover, the fact that the interface that we select is not the perfect one, in that there are grains present on either side of the interface that should not be there, is taken into account when we compute the segregation index through \eqref{eq:index}, which contrasts the observed segregated state with the ideal state that we define through our choice of the interface.

\begin{figure}
\centering
\includegraphics[scale=0.5]{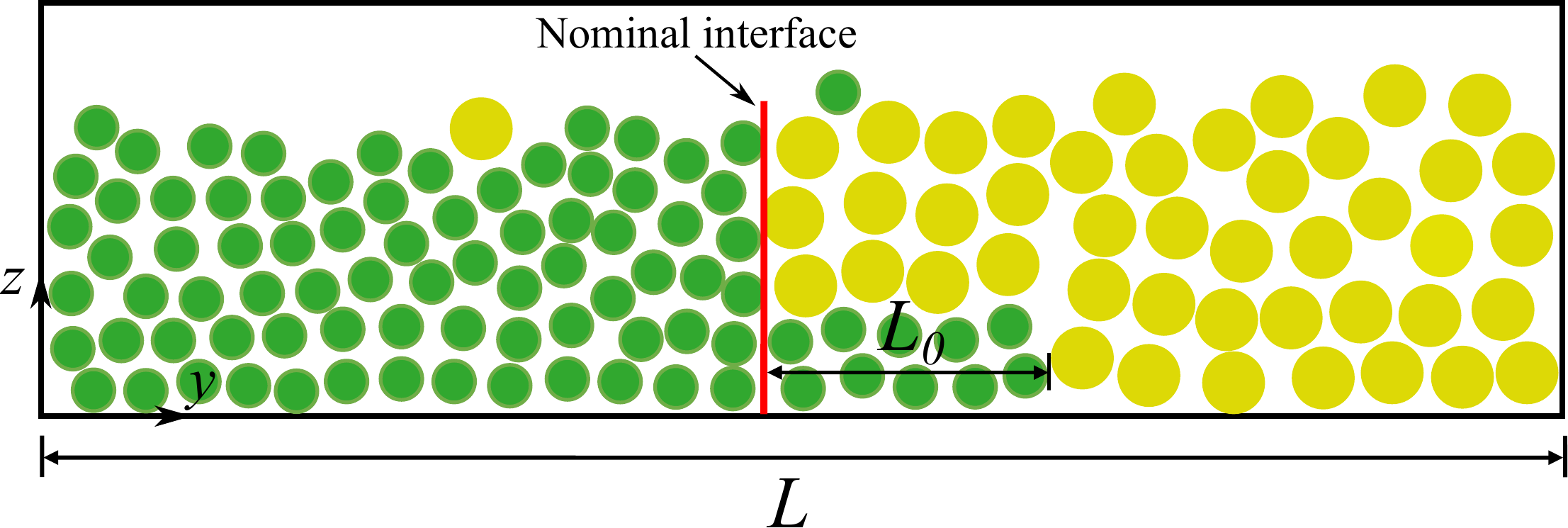}
\caption{A schematic illustrating the overlap length $L_0$ in a mixture of small green grains and large yellow grains. Here, the side view along the shaking direction is shown.}
\label{fig:overlap}
\end{figure}

We note that the computation of the segregation index utilizes the top view of the system, which may not provide us with a complete picture about the extent of segregation through the depth of the mixture bed. Thus, we simultaneously examined the bulk state of the mixture by calculating the degree of overlap of the two grain species near the interface in order to obtain through-depth information. This overlap reflects the distance through which the small grains may have diffused into the zone nominally identified as being comprised of only big grains. For measuring this overlap, the top layer of big grains was manually probed by gently brushing aside the surface grains at the center of the channel along the axis to check for the presence of small grains underneath. We measured the shortest distance from the nominal interface -- as identified above to define the ideally segregated system -- to the point lying along the channel's axis, to the right of the nominal interface, beyond which no small grains were found, as shown in the schematic of Fig.~\ref{fig:overlap}. This overlap length ($L_O$) is then normalized by the channel's length, to define the {\em normalized overlap} as
\begin{equation}
\label{eq:no}
\lambda=\frac{L_O}{L}.
\end{equation}
Note that a lower $\lambda$ corresponds to better segregation.

Here we rely upon manual probing because our granular bed is three-dimensional and opaque, and comprises of a large number of grains. The opacity of granular materials and  related issues in investigating  through-depth  dynamics of granular systems have been discussed in detail by Amon \textit{et al.} \cite{amon2017}. We note that although our defined normalized overlap $\lambda$ is less constrained compared to the segregation index $\sigma$, its correlation with $\sigma$, if established, will be useful and will lend confidence to our quantification of the state of segregation.

\section{Results: Experiments}
\label{sec:results-E}
\subsection{Effect of size ratio and frequency on segregation}

\begin{figure}[b!]
\centering
\includegraphics[scale=0.6]{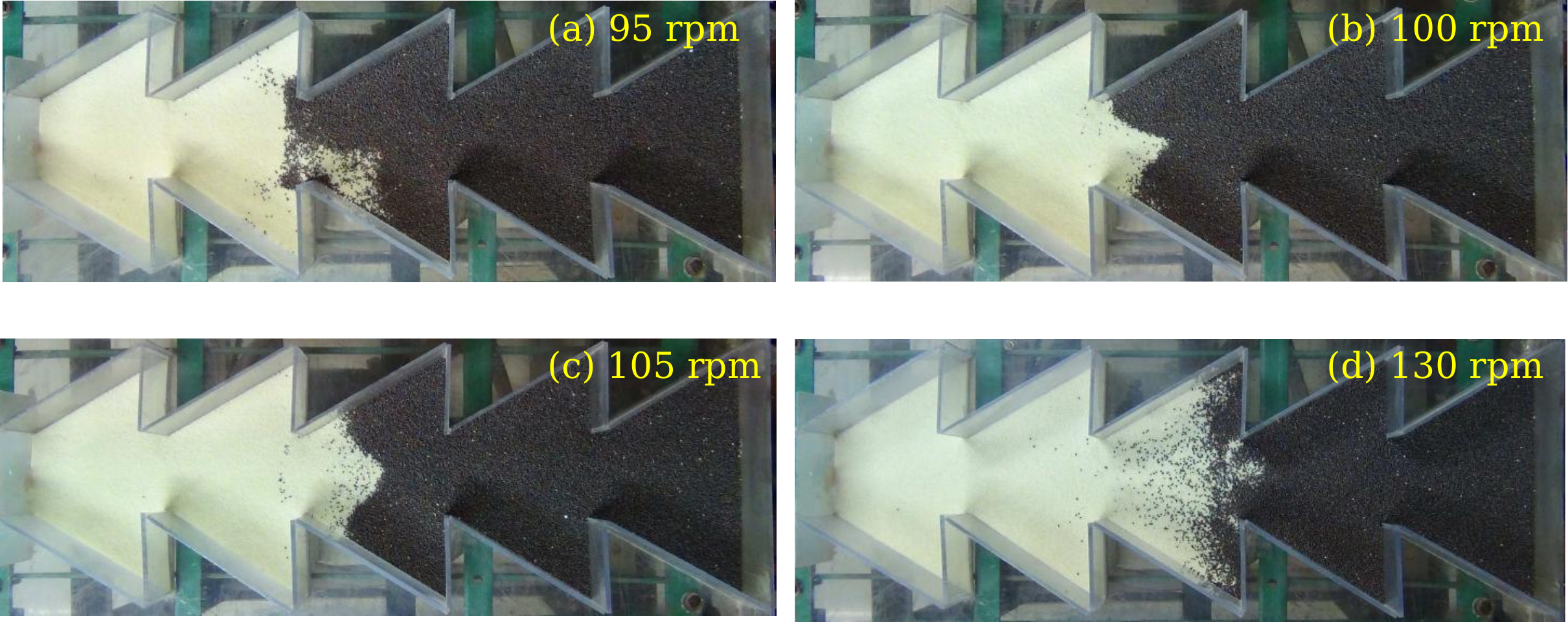}
\caption{Final segregated states of the millet and black-mustard mixture shaken in the $60^{\circ}$ taper channel at four different frequencies whose values are provided in the images. Visually, the best segregation was observed at $105$ \text{rpm}.}
\label{fig:MM60}
\end{figure}

We now investigate the quality of segregation in binary mixtures of foodgrains experimentally. Table~\ref{tab:table2} lists the various mixtures that we consider. Recall that the density ratio $\rho_r\approx1$ for all these mixtures. To begin our discussion, we consider experiments conducted on a binary mixture of millet and black mustard grains with size ratio $s_r=1.47$ and density ratio $\rho_r=1.125$, confined in a regular multi-trapezium channel with taper angle $\alpha=60^{\circ}$. The channel is filled with the mixture up to an initial height of $17$ small grain (millet) diameter. The images of the final segregated states at several frequencies are displayed in Fig.~\ref{fig:MM60}.
We  did not observe any discernible segregation for this mixture below 95 \text{rpm} as the grains collectively move with the channel like a solid body. At 95 \text{rpm}, segregation may be clearly observed, although it is of poor quality; see Fig.~\ref{fig:MM60}(a). Further increasing the shaking frequency improves the segregation quality until $N=105$ \text{rpm} -- see Figs.~\ref{fig:MM60}(b) and (c) --  beyond which the quality of segregation is  found to decrease as depicted in Fig.~\ref{fig:MM60}(d). It is also evident from Figs.~\ref{fig:MM60}(b) and (c) that the interface between the two species is very sharp and nearly V-shaped at $N=100$ \text{rpm} and $105$ \text{rpm}. As $N$ increases further, the final segregated states show  mustard  grains spread in the vicinity of the interface, thereby making it dull, even as the interface assumes a nearly straight shape. This is observed, for example, in Fig.~\ref{fig:MM60}(d), which corresponds to the highest shaking frequency $N=130$ \text{rpm} that we report.

\begin{figure}
\centering
\includegraphics[scale=0.65]{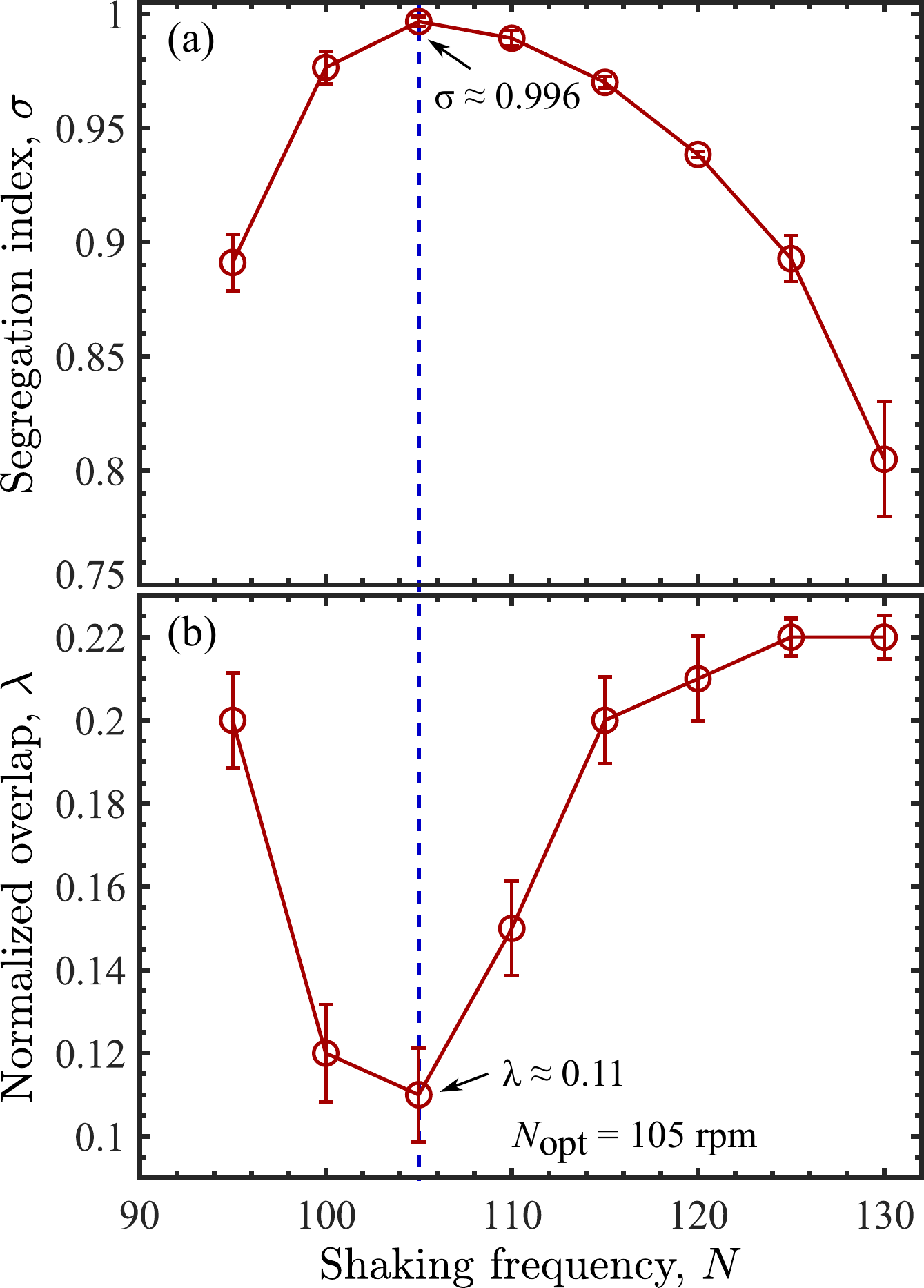}
\caption{Variation of the (a) segregation index $\sigma$ and (b) normalized overlap $\lambda$ with  frequency $N$ for the  mixture of millet and mustard shaken in the $60^\circ$ taper channel; cf. Fig.~\ref{fig:MM60}. Error bars represent the standard deviation computed over five data points for each frequency.}
\label{fig:MM60_SigLam}
\end{figure}

The above aspects are illustrated quantitatively in Fig.~\ref{fig:MM60_SigLam}, which displays the variations of the segregation index $\sigma$ and the normalized overlap $\lambda$ with the shaking frequency $N$. We observe that, as the shaking frequency is raised, the parameter $\sigma$  increases and reaches a maximum before decreasing, while  $\lambda$  follows an opposite trend. Recalling that lowering of $\lambda$ indicates better separation of the species through the mixture's depth, this confirms that the overall segregation quality -- surface and through-depth -- improves initially when the system is shaken more vigorously, but diminishes at high shaking frequencies. Very interestingly, Fig.~\ref{fig:MM60_SigLam} shows that the maximum  segregation index, $\sigma_{\max} \approx 0.996$, occurs at the shaking frequency $N_{\text{opt}} = 105$ \text{rpm}, which is also the frequency at which the normalized overlap attains a minimum, $\lambda_{\min} \approx 0.11$. Although this feature is not true for all mixtures which we investigate, the frequencies at which $\sigma_{\max}$ and $\lambda_{\min}$ are achieved are typically close together. Thus, this indicates that the best segregation is achieved over a {\textit{unique}} narrow range of optimal frequencies.  Finally, Fig.~\ref{fig:MM60_SigLam} indicates that the overlap grows  as $N$ increases beyond 105 \text{rpm} and saturates to a maximum value of $22\%$. The segregation index $\sigma$ too decays gradually for $N > 105$ \text{rpm}. The high value of $\sigma$ and the low $\lambda$ at 130 \text{rpm} corroborate the poor quality of the observed segregation.

\begin{figure}
\centering
\includegraphics[scale=0.6]{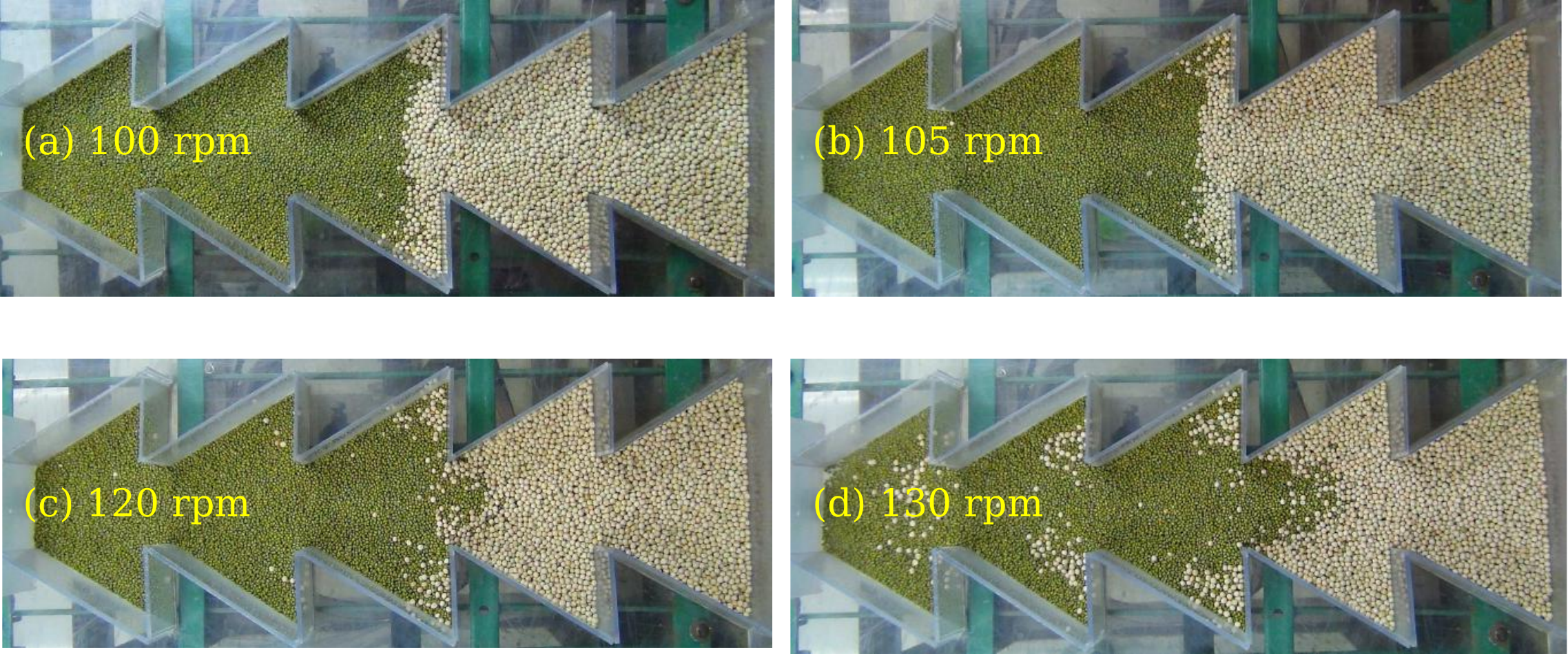}
\caption{Final segregated states of the peas and green-gram mixture shaken in the $60^{\circ}$ taper channel at different frequencies whose values are provided in the images. Visually, the best segregation was observed at $105\ \text{rpm}$.}
\label{fig:PGG60}
\end{figure}
\begin{figure}
\centering
\includegraphics[scale=0.65]{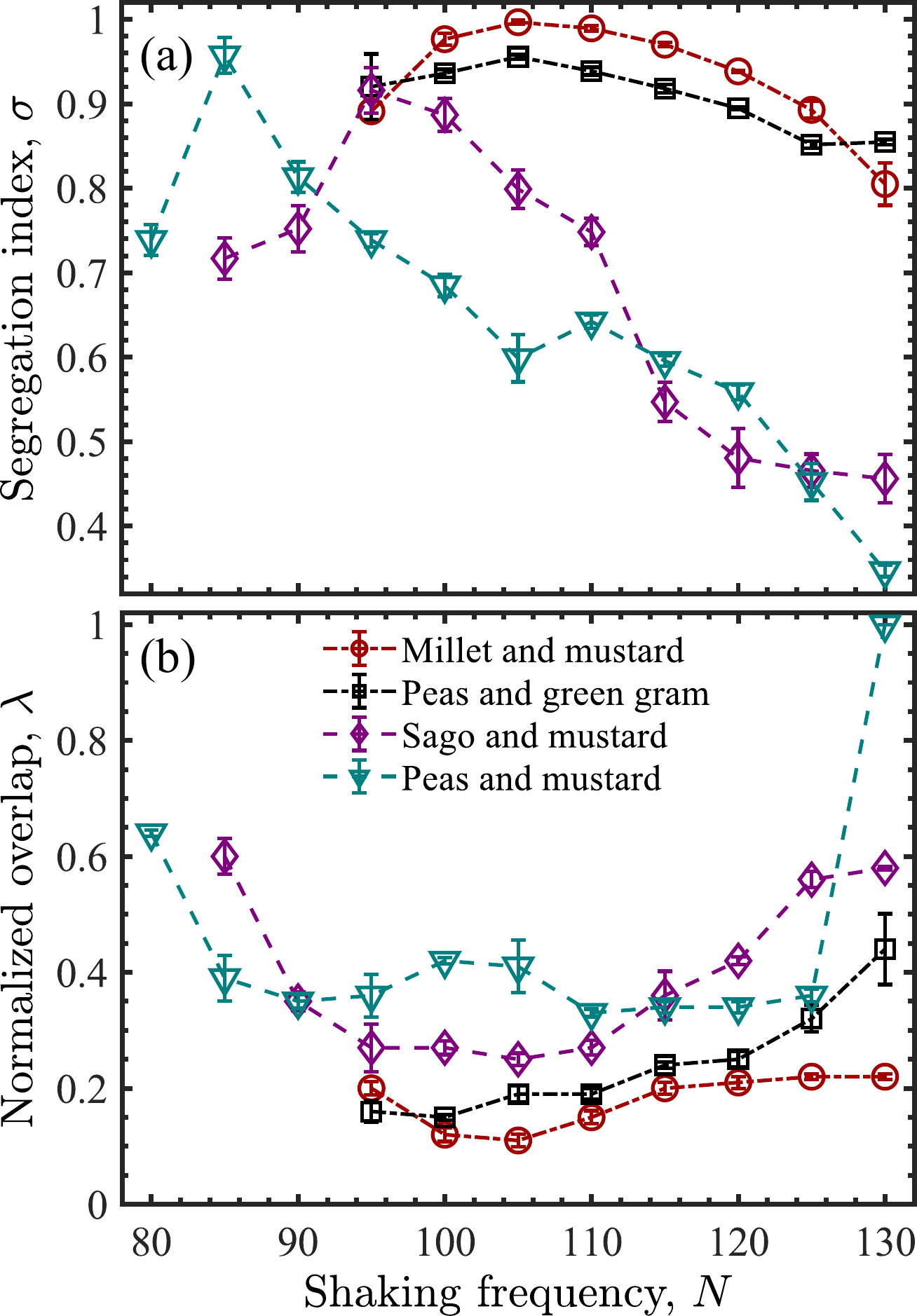}
\caption{Variation of the (a) segregation index $\sigma$ and (b) normalized overlap $\lambda$ with  frequency $N$ for four binary mixtures shaken in the $\alpha=60^\circ$ taper channel. Error bars denote the standard deviation computed over five data points for each frequency. Legend for both plots is given in (b).}
\label{fig:All_SigLam}
\end{figure}

The second mixture that we present here is of peas and green gram, having the size ratio $s_r=1.4$ and an initial filling height of $10$ small grain size (minimum diameter of green gram). The final segregated states at several frequencies are shown in Fig.~\ref{fig:PGG60}. We notice that the  interface between the two species at low frequencies is nearly straight, which transforms to a curvilinear shape at frequencies equal and higher than 120 \text{rpm}. The quality of segregation is poorer in comparison to the millet-mustard mixture discussed above, but still reasonably good. This feature is reflected in Fig.~\ref{fig:All_SigLam} that plots the segregation index $\sigma$ and the normalized overlap $\lambda$ for all the mixtures that we report here. We observe from Fig.~\ref{fig:All_SigLam}(a) that the curve for the segregation index for the peas and green gram mixture lies below the corresponding curve for the millet and mustard mixture.  Similarly, in Fig.~\ref{fig:All_SigLam}(b), the curve for the normalized overlap $\lambda$  lies above the  one for the millet and mustard mixture, indicating longer axial diffusion of green gram into the zone populated by peas. Moreover, as displayed in Fig.~\ref{fig:All_SigLam}, the frequency corresponding to the greatest segregation index, $N^\sigma_{\text{opt}}\sim105\ \text{rpm}$, does not yield the best segregation through the depth of the mixture, which occurs when $N^\lambda_{\text{opt}}\sim100\ \text{rpm}$. However, the $\lambda$ curve is reasonably flat, so that through-depth segregation remains nearly optimal even when only $\sigma$ is maximized.

\begin{figure}
\centering
\includegraphics[scale=0.6]{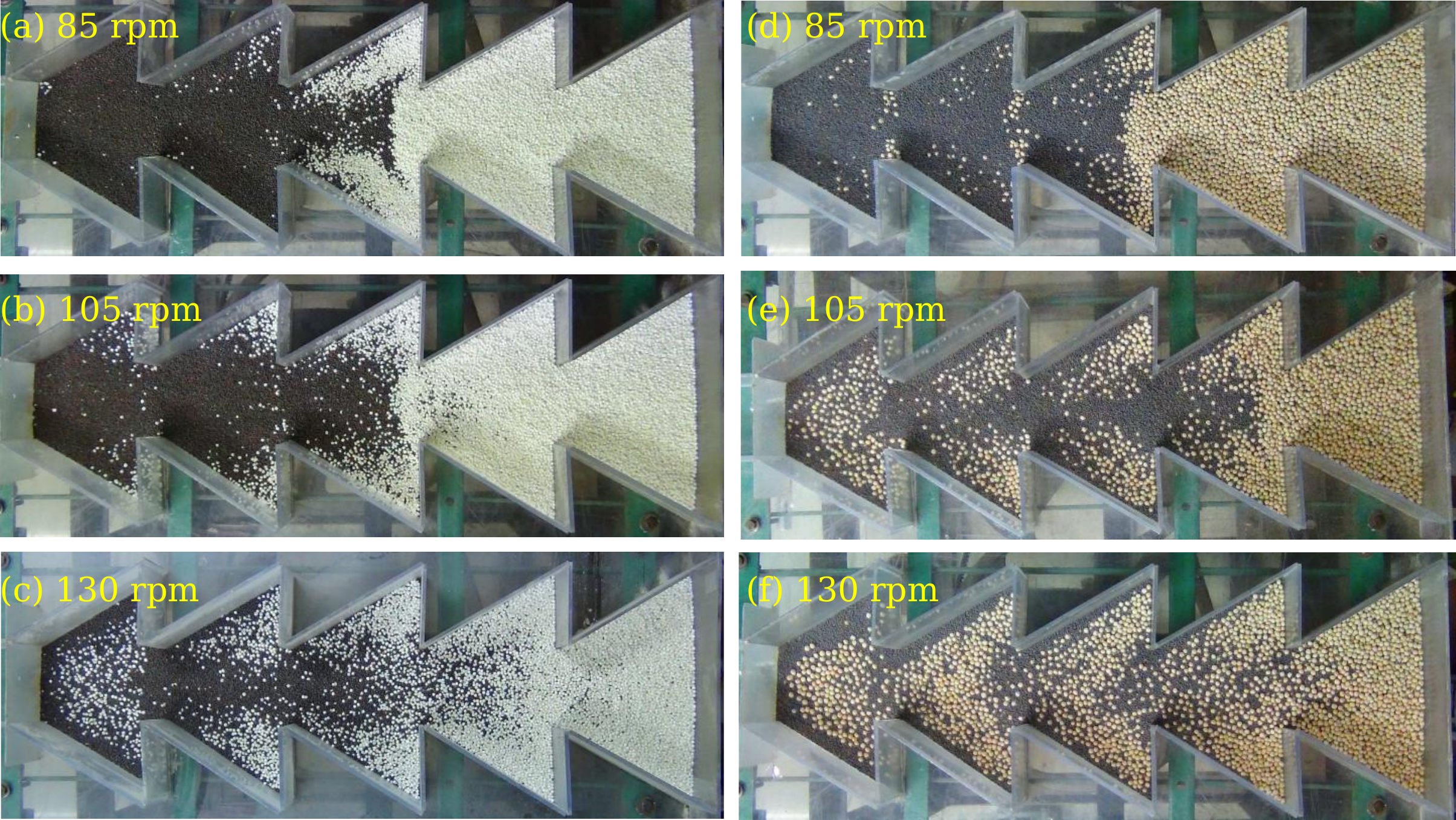}
\caption{Final segregated states of  (a)-(c) sago and mustard  and (d)-(f) peas and mustard mixtures. The mixtures are shaken in the $60^{\circ}$ taper channel at different frequencies whose values are provided in the images. Visually, the best segregation was observed at $95$ \text{rpm} for the sago and mustard mixture and at $85$ \text{rpm} for the peas and mustard mixture.}
\label{fig:spm60}
\end{figure}

We next discuss segregation in a mixture of (white) sago and (black) mustard grains that have size ratio $s_r=1.86$ and an initial filling height of $17$ small grain (mustard) diameter. We observe that the quality of segregation is relatively poor in comparison to the previous two mixtures, as shown in Figs.~\ref{fig:spm60}(a)-(c). This  is quantitatively evident from Fig.~\ref{fig:All_SigLam}(a) where the curve for the segregation index  for the sago-mustard mixture lies below the $\sigma-$curve for the millet-mustard and peas-green gram mixtures. Moreover, the frequency at which maximum segregation $\sigma_{\max}$ occurs, again, does not lead exactly to minimum of normalized overlap $\lambda_{\min}$. However, as for the peas-green gram mixture, the $\lambda$-curve is flat bottomed, so that, again, through-depth segregation remains close to its optimal value at the frequency that maximizes $\sigma$. In passing, we observe that some sago grains remain trapped at the corners of the trapezoidal section due to the development of dead zones. These zones may be relieved by blunting the sharpness of those corners of the trapezium. 

Finally, we consider the mixture of (off-white) peas and (black) mustard grains with size ratio $s_r=3.18$ and an initial filling height of $15$ small grain (mustard) diameter; see Figs.~\ref{fig:spm60}(d)-(f). We observe that the quality of segregation is similar to that of the sago-mustard mixture. However, as Fig.~\ref{fig:All_SigLam} shows, the highest value for the segregation index $\sigma$ is achieved at a frequency of $N=85$ \text{rpm} in this mixture, which is lower than the previous three mixtures. We find  a sharply peaked response for $\sigma$, suggesting sensitivity of the segregation to the shaking frequency $N$. However, the overlap $\lambda$ continues to display a flat-bottomed response, as seen in Fig.~\ref{fig:All_SigLam}(b). Thus, even though the frequency at which $\sigma_{\max}$ occurs does not match the frequency corresponding to $\lambda_{\min}$, one may still get near-optimal through-depth segregation when $\sigma$ is maximized. Furthermore, we found a $100\%$ overlap at $130$ \text{rpm}, at which point $\sigma$ is also very low, indicating no segregation. This is in contrast to the other mixtures where some segregation, albeit poor, was  observed even at 130 \text{rpm}. Lastly, this mixture also exhibits the effect of dead zones that, here, grow and meet each other.

The plots of Fig.~\ref{fig:All_SigLam} allow us to make the following general observations about the segregation in the four mixtures that we investigated. First, although millet-mustard mixture has higher values of segregation index $\sigma$ than the peas-green gram mixture, the $\sigma$ value of peas-green gram  mixture surpasses that of the millet-mustard mixture at the two ends of the frequency range, i.e. at $N=95$ \text{rpm} and $130$ \text{rpm}. Similarly, at $N=95$ \text{rpm} the $\lambda$ value of millet-mustard mixture exceeds that of the peas-green gram mixture. Thus, at the minimum operating frequency of $N=95$ \text{rpm}, the peas-green gram mixture segregates better than the millet-mustard mixture. Also, the segregation index curves in Fig.~\ref{fig:All_SigLam} for both peas-mustard and sago-mustard grains mixtures have sharper peaks compared to the curves for the millet-mustard and peas-green gram mixtures. This suggests that segregation in millet-mustard and peas-green gram mixtures is less sensitive to shaking frequency. Moreover, the fall in the quality of segregation for peas-mustard and sago-mustard mixtures is very rapid, and the minimum value of segregation index $\sigma_{\min}$ is much lower than that for millet-mustard and peas-green gram mixtures. This indicates that at higher frequencies the former two mixtures exhibit poorer segregation compared to the latter two mixtures, as is also evident when we compare  Fig.~\ref{fig:spm60} with Figs.~\ref{fig:MM60} and \ref{fig:PGG60}.

We now proceed to investigate the effect of various system parameters on the segregation.

\subsection{Effect of channel taper ($\alpha$)}
\begin{figure}
\centering
\includegraphics[scale=0.65]{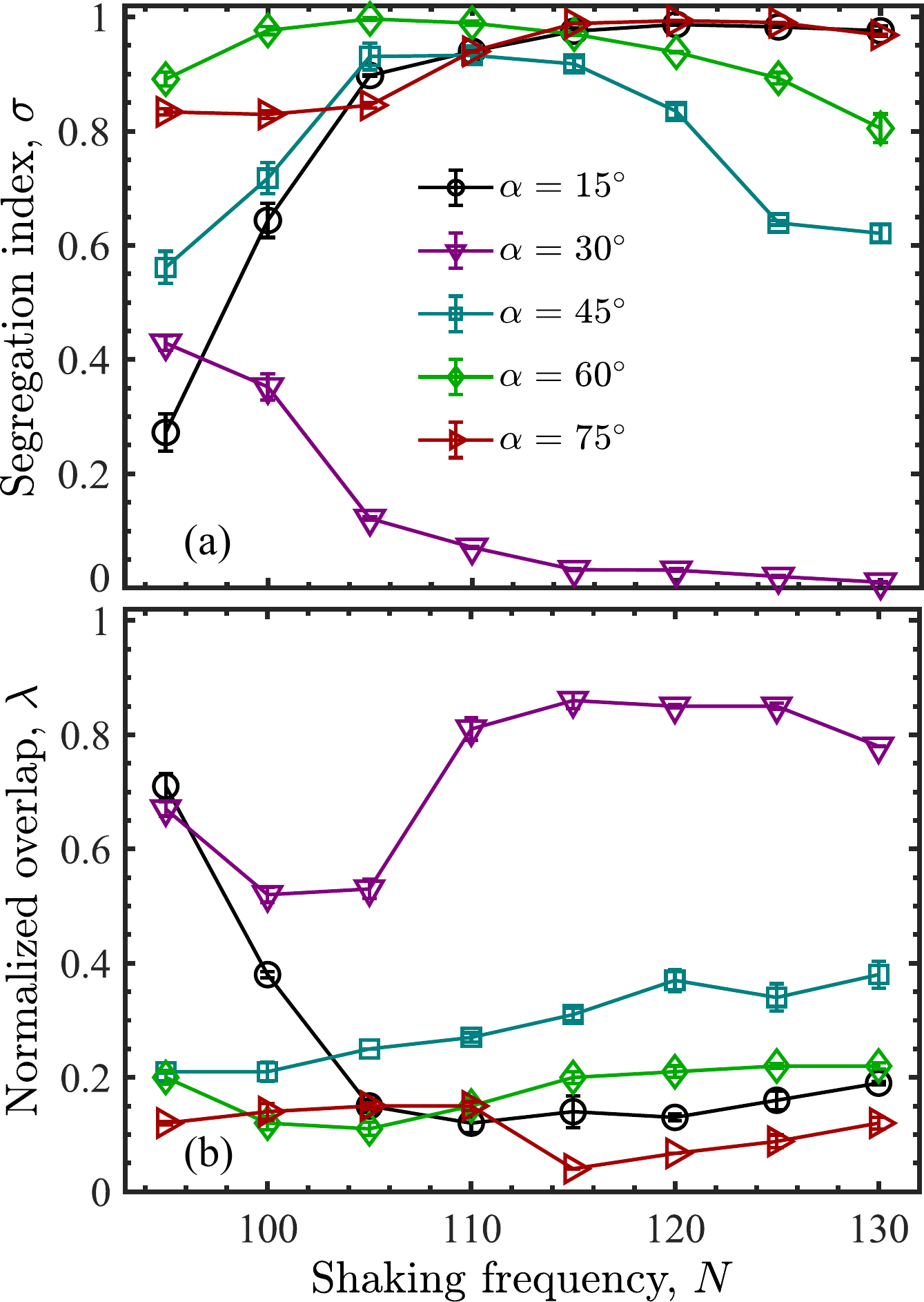}
\caption{Variation of the (a) segregation index $\sigma$ and (b) normalized overlap $\lambda$ with frequency $N$ for the binary mixture of millet and mustard for several  taper angles $\alpha$. Error bars denote the standard deviation computed over five data points. Legend for both plots is provided in (a).}
\label{fig:MM_alpha_SigLam}
\end{figure}

We begin by  discussing the effect of channel taper $\alpha$ on the quality of segregation. We varied the taper angle of channel's sidewalls and conducted experiments with millet and mustard mixture, which we saw above to display, arguably, the best segregation in a $\alpha = 60^\circ$ channel. Note that the number of sections changes while varying $\alpha$, keeping the dimensions $l_e$, $u_e$ and $L$ same. Figure~\ref{fig:MM_alpha_SigLam} displays the plots for the segregation index $\sigma$ and normalized overlap $\lambda$  for a millet-mustard mixture shaken in differently tapered channels. We observed that the channels with taper angles $\alpha=45^{\circ}, 60^\circ$ and $75^{\circ}$ yield comparable segregation outcomes, which is corroborated by the similarity in the behaviors of the $\sigma$ and $\lambda$ curves. In contrast, the $\alpha=15^{\circ}$ and $30^{\circ}$ channels exhibit anomalous and weak segregation, respectively, and we discuss them separately further below. 

\begin{figure}
\centering
\includegraphics[scale=0.5]{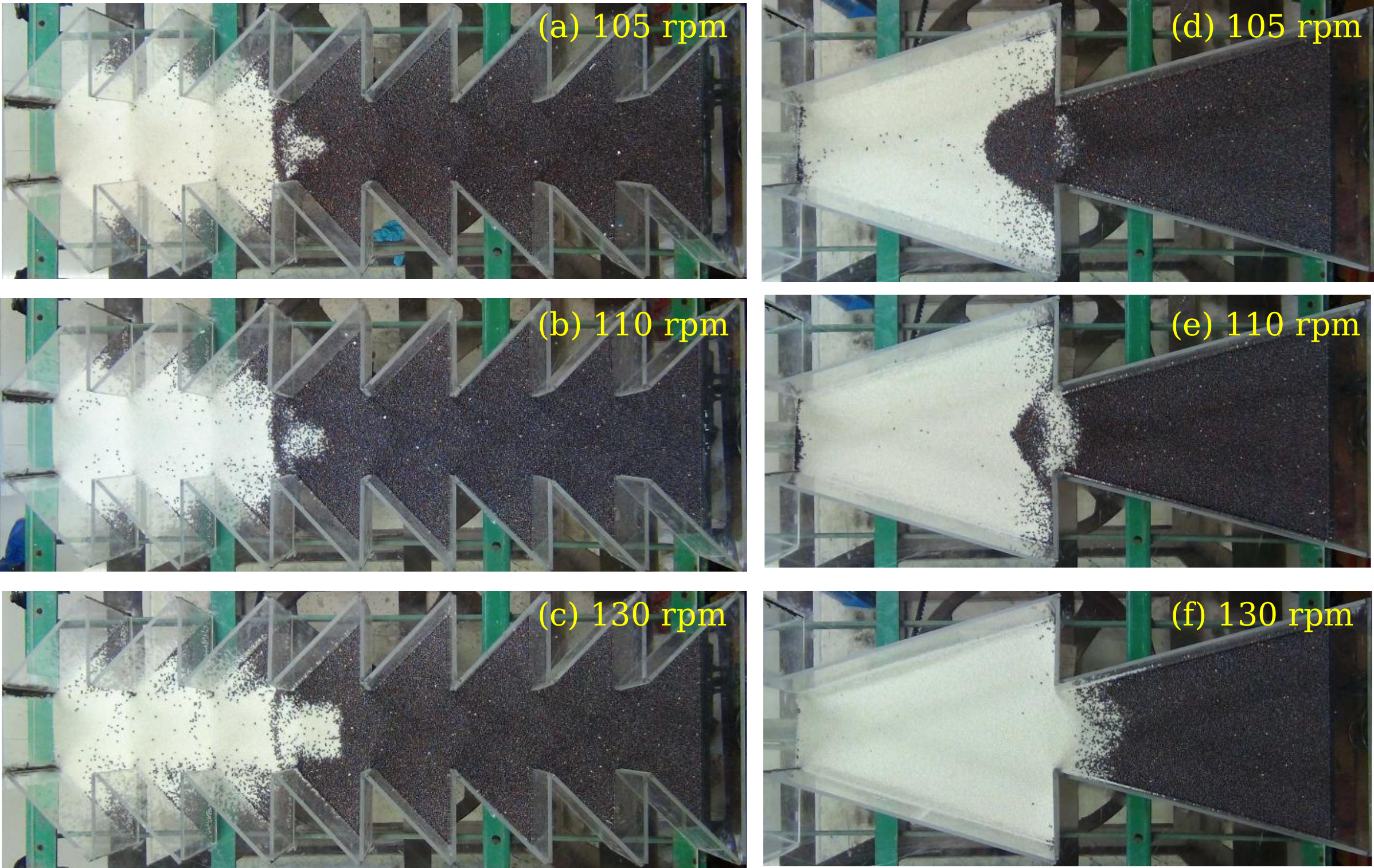}
\caption{Final segregated states of the millet and mustard mixture shaken in (a)-(c) a $45^{\circ}$ taper channel and (d)-(f) a $75^{\circ}$ taper channel at three different frequencies whose values are provided in the images. Visually, the best segregation is observed at $110$ \text{rpm} in both channels.}
\label{fig:MM45_75}
\end{figure}

The segregation index $\sigma$ for the $60^{\circ}$ taper channel was discussed in the preceding section. The $\sigma$ and $\lambda$ curves for this channel are almost flat, displaying good segregation quality at  frequencies $N$  around the optimum value $N_{\text{opt}}$ at which $\sigma_{\max}$ is achieved.  The $\sigma$ and $\lambda$ curves for the $45^{\circ}$ taper channel lie, respectively, below and above the corresponding curves for  the  $60^{\circ}$ taper channel, signifying comparatively poorer segregation in the former channel. Nevertheless, the peak segregation index in the $45^{\circ}$ is still higher than $90\%$, with the corresponding $\lambda \lessapprox 20\%$. Figures~\ref{fig:MM45_75}(a)-(c) display images of the final segregation in a $45^\circ$ channel at three different $N$. From the images of Figs.~\ref{fig:MM45_75}(a)-(c) we may also explain the reason for  the  poorer performance of the $45^\circ$ as follows: the segregation quality is lowered by the formation of dead zones whose tally and extent are  augmented by the increased number of trapezia when the taper is reduced to $45^\circ$ from $60^\circ$.

Turning to the $75^\circ$ channel, Fig.~\ref{fig:MM_alpha_SigLam} shows that the segregation index curve for the $75^\circ$ channel lies above the one for the $45^\circ$ channel, except for a narrow range of frequencies around $N = 105$ \text{rpm}. Interestingly, the $\sigma$ for the segregation achieved in the $75^\circ$ channel is better than  the $60^\circ$ channel for $N \geqslant 115$ \text{rpm}, but poorer for smaller $N$.    The normalized overlap  $\lambda$ for the $75^\circ$ channel behaves analogously. We observe that the $75^\circ$ channel needs to be shaken at a higher frequency in order to achieve the same level of segregation as the $60^\circ$ channel. We explain this as follows. When the taper is large, the channel has to be shaken more vigorously in order to provide the required level of  axial momentum to drive the mixture to the channel's right end, which is a necessary first step in the segregation process. Figures~\ref{fig:MM45_75}(d)-(f) display images of the final segregation in a $75^\circ$ channel at three different $N$. We note the reduced number and the lower  extent of the dead zones.

\begin{figure}
\centering
\includegraphics[scale=0.6]{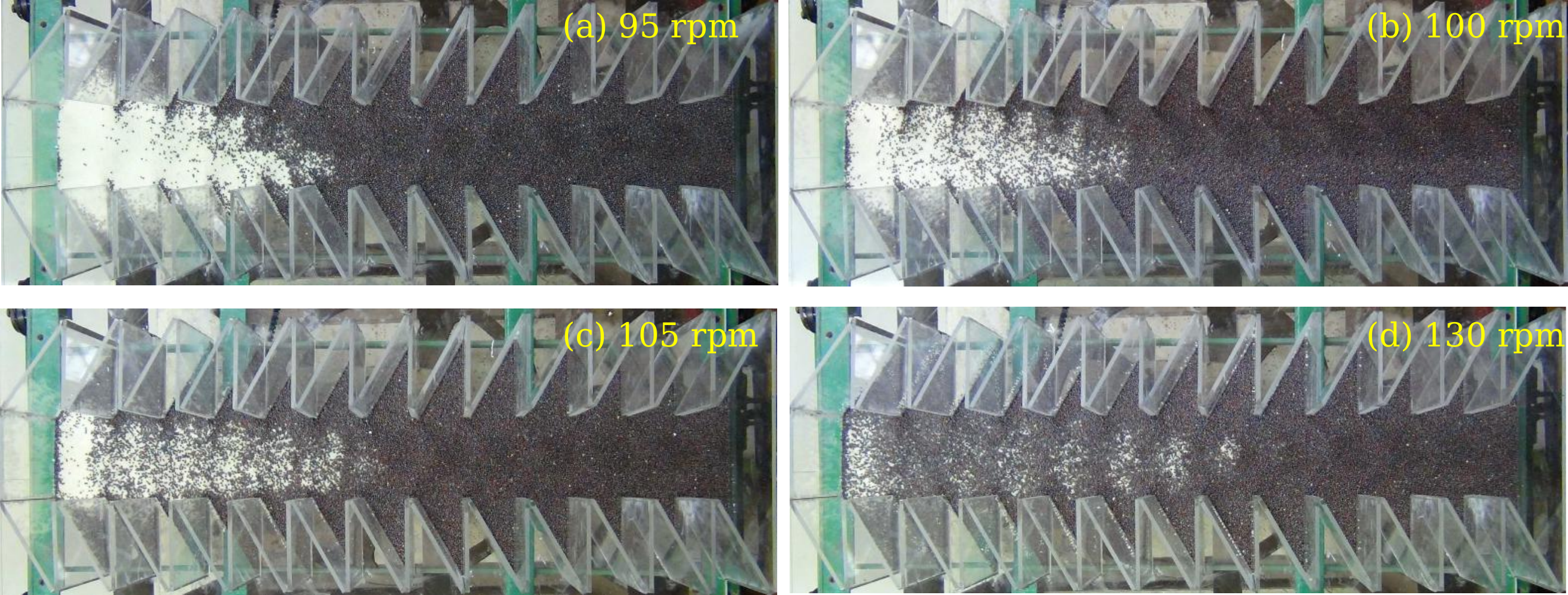}
\caption{Final segregated states of the millet and mustard mixture shaken in the $30^{\circ}$ taper channel at different frequencies whose values are provided in the images. Visually, the best segregation is observed at $95$ \text{rpm}.}
\label{fig:MM30}
\end{figure}

 
We turn now to the weak and anomalous segregation outcomes that we observed in the $\alpha = 30^\circ$ and $15^\circ$ channels, respectively. The $30^{\circ}$ taper channel displays weak segregation at $95$ \text{rpm} and $100$ \text{rpm}, and almost no segregation as we further increase the frequency, which is evident from Fig.~\ref{fig:MM30}. This aspect is also reflected quantitatively in the plots of the segregation index and normalized overlap  in Fig.~\ref{fig:MM_alpha_SigLam}, where $\sigma \lessapprox 0.4$ and $\lambda \gtrapprox 50\%$. We note that, while vertical segregation takes place effectively, with the larger mustard seeds lying on top of the smaller millet grains, the axial separation of the grains is largely impeded by the incomplete accumulation of the mustard seeds at the right end of the channel. This, we believe,  is caused by the presence of many narrow trapezia in a $30^\circ$ channel that increase the number and extent of the dead zones which, in turn, obstruct the rightward motion of grains. 

\begin{figure}
\centering
\includegraphics[scale=0.7]{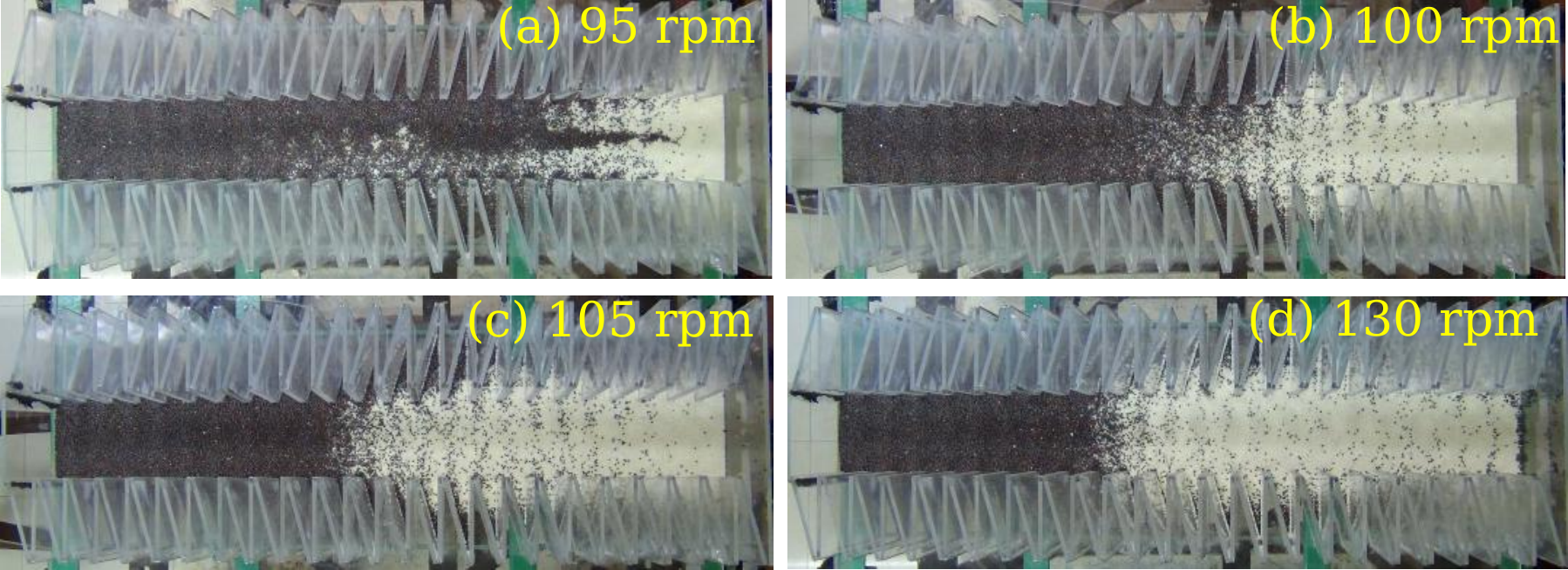}
\caption{Final segregated states of the millet and mustard mixture shaken in the $15^{\circ}$ taper channel at different frequencies whose values are provided in the images. Visually, the best segregation is observed at $120$ \text{rpm} (not shown).}
\label{fig:MM15}
\end{figure}

We expected outcomes similar to the $30^\circ$ taper channel from a $15^\circ$ channel. We were, however, greatly surprised to find that the $15^{\circ}$ taper channel shows {\em reverse} segregation, wherein the small grains (millet) accumulate near the right end of the channel and big grains (mustard) accumulate at the left end of the channel, as displayed in Fig.~\ref{fig:MM15}. The quality of the segregation is also rather good at shaking frequencies $N\gtrapprox 105$ \text{rpm}, as confirmed by the high/low values of  segregation index $\sigma$/ normalized overlap $\lambda$ in Fig.~\ref{fig:MM_alpha_SigLam}. At low $N$, $\sigma$ is less than $0.65$ and the corresponding values of $\lambda$ are more than $15\%$, which correspond to poor  segregation that  is also noticeable in Figs.~\ref{fig:MM15}(a) and (b). At present we are unable to explain this observation completely, but hypothesize that the reason may lie in the manner the many dead zones that are formed in a $15^\circ$ channel extend and interact. This remains a problem to investigate in the future. 

\begin{figure}
\centering
\includegraphics[scale=0.5]{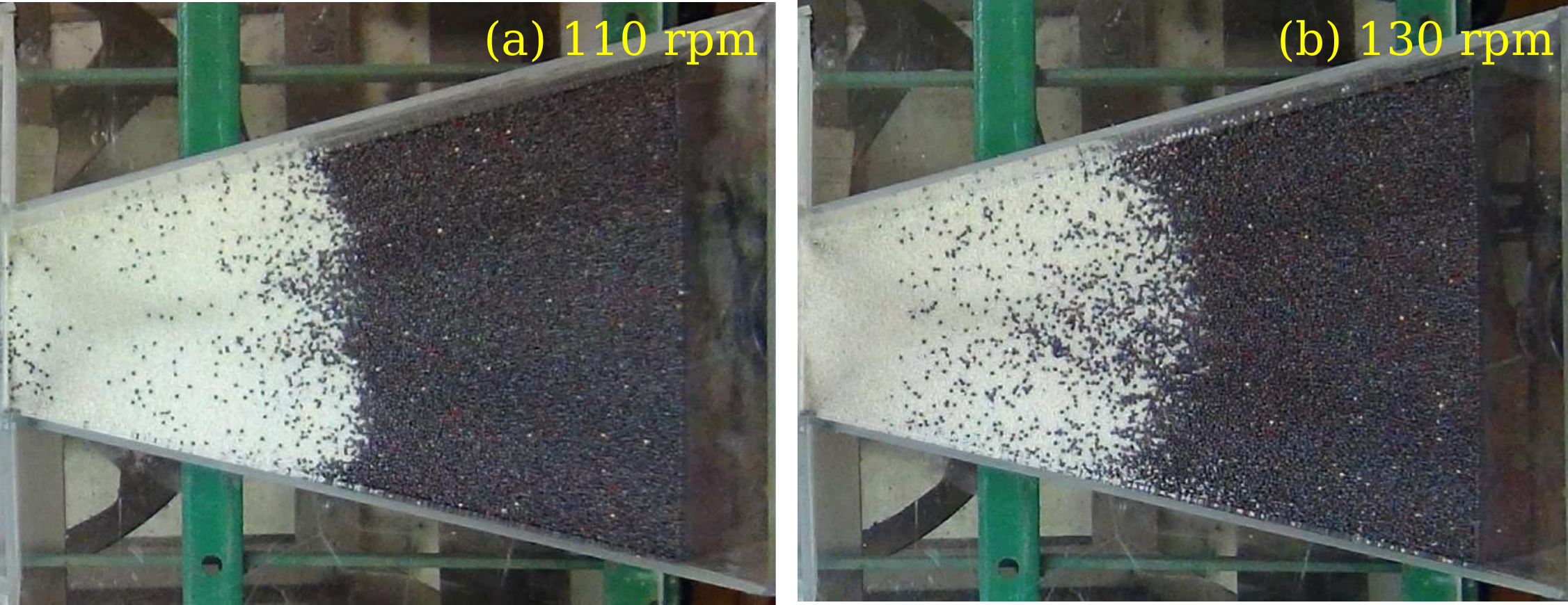}
\caption{Final segregated states of the millet and mustard mixture shaken in the $75^{\circ}$ taper channel, having only one trapezium section, at different frequencies mentioned on each figure. The best segregation is achieved at $125$ \text{rpm}.}
\label{fig:MM75_OneSec}
\end{figure}

We also conducted experiments with millet and mustard mixture confined in a $75^{\circ}$ single trapezium channel. Figure~\ref{fig:MM75_OneSec} displays the final segregated states and we note that this channel is also able to segregate the mixture reasonably well. Thus, we conclude that concatenation of the trapezium sections is not required to achieve segregation, rather only tuning the channel taper is needed to achieve better segregation. This single trapezium channel is simple in design, avoids unnecessary complications in fabricating the multi-trapezium channel, and thus could be utilized as  a surrogate for the multi-trapezium channel.

\subsection{Effect of channel offset ($l_o$)}
\begin{figure}
\centering
\includegraphics[scale=0.32]{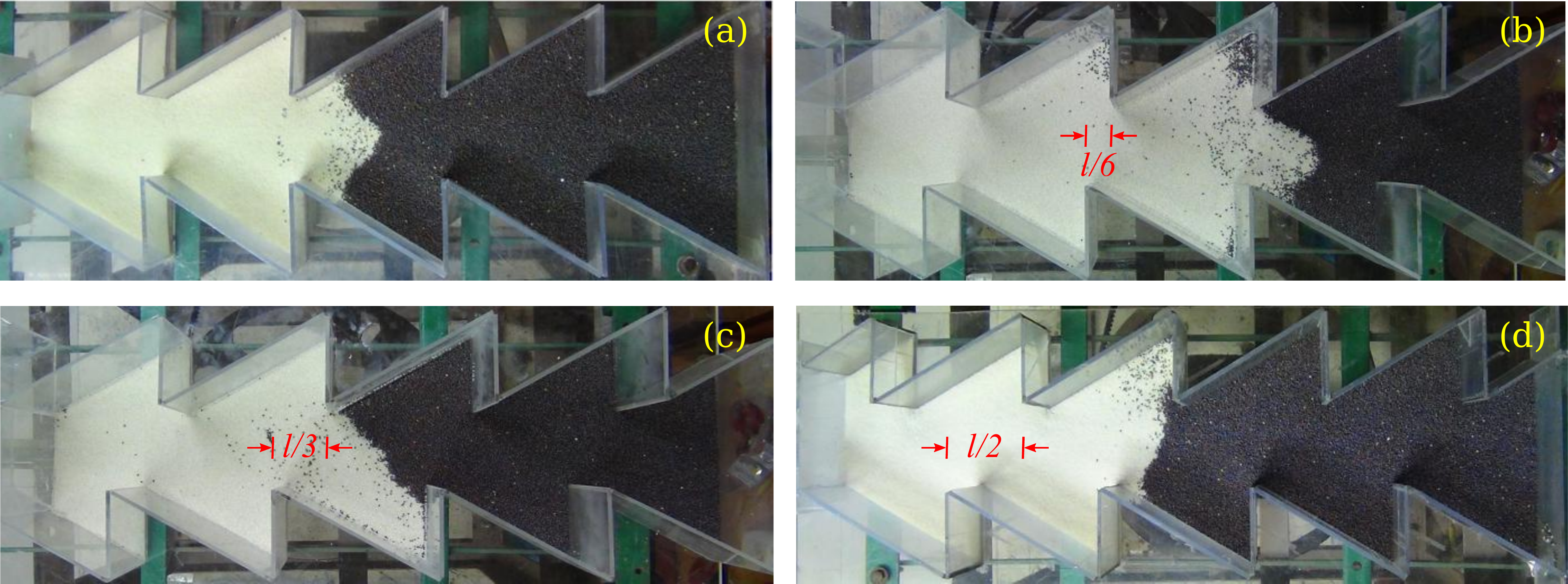}
\caption{Comparison of the  segregated states achieved at the optimum frequency $N_{opt}=105\ \text{rpm}$ for the millet and mustard mixture shaken in the $60^{\circ}$ taper channel with several offsets $l_o$: (a) no offset $(l_o= 0)$, (b) $l_o= l/6$, (c) $l_o = l/3$, and (d) $l_o = l/2$.}
\label{fig:MM60_offset}
\end{figure}

So far we focussed on channels with no offset, i.e. $l_o = 0$; cf. Fig.~\ref{fig:setup}. We now investigate the effect of staggering or offsetting the trapezoidal sections on the quality of segregation  in $60^{\circ}$ taper channel. We hypothesize that offsetting will improve segregation, as it will  help release the dead zones that form in the corners of the trapezia. Figure~\ref{fig:MM60_offset} compares the final segregated state in a channel with no offset with those obtained in staggered channels with offsets $l_o = l/6, l/3$ and $l/2$, where we recall that  $l$ is the axial length of a trapezoidal section in a  multi-trapezium channel when $l_o = 0$. Note that, because of translational periodicity, there is no need to study channels with offset greater than $l/2$. As in the preceding section, we employ a millet-mustard mixture.

\begin{figure}
\centering
\includegraphics[scale=0.65]{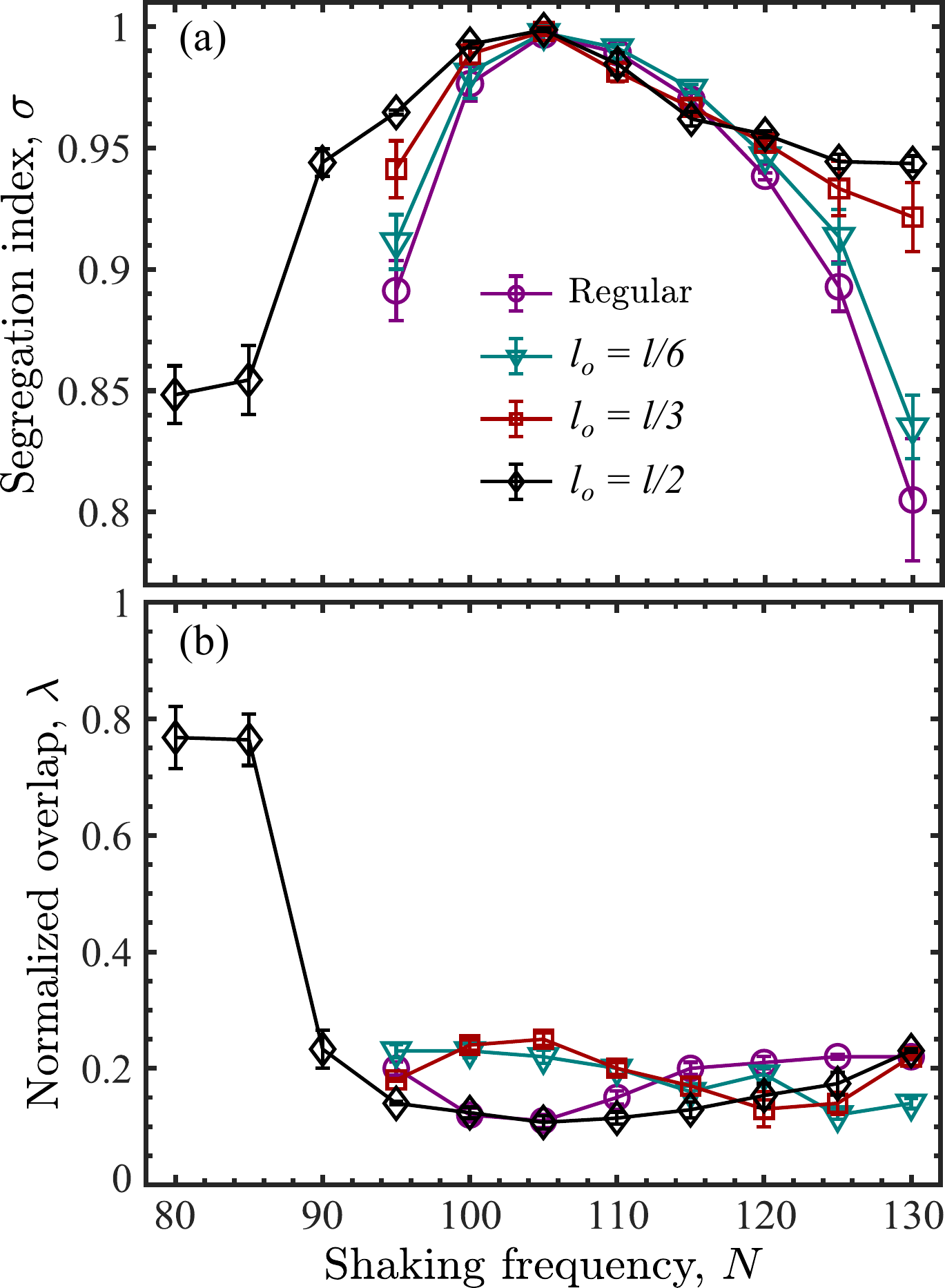}
\caption{Variation of the (a) segregation-index ($\sigma$) and (b) normalized overlap ($\lambda$) with frequency $N$ for the millet and mustard mixture shaken in the regular and staggered channels with various offsets $l_o$. Error bars denote the standard deviation computed over five data points. Legend for both plots is provided in (a).}
\label{fig:MM_offset_SigLam}
\end{figure}

Figure~\ref{fig:MM_offset_SigLam} depicts the variation of segregation index $\sigma$ and the normalized overlap $\lambda$ with the shaking frequency $N$ for the four  channels mentioned above. We find that the best segregation is achieved in the channel with the  largest offset $l_o = l/2$, and the segregation quality decreases with offset, so that the channel with no offset ($l_o = 0$) performs relatively most poorly. The optimum frequency to obtain the best segregation remains $N_{\text{opt}}=105$ \text{rpm}, as is evident from Fig.~\ref{fig:MM_offset_SigLam}(a). We do not observe much difference in $\sigma$ for these four cases near $N_{\text{opt}}$. However, providing an offset flattens the $\sigma$ curve, so that segregation quality remains high over a much larger range of shaking frequencies, as is clear from Fig.~\ref{fig:MM_offset_SigLam}(a). Indeed, the $l/2-$offset channel exhibits very good segregation even at the low $N$ of  90 \text{rpm} and the high $N$ of 130 \text{rpm}.  Below $90$ \text{rpm}, while $\sigma$ remains acceptably high,  the overlap in the $l/2-$offset channel is too great -- see  Fig.~\ref{fig:MM_offset_SigLam}(b) --  so that the overall segregation quality lowers. In contrast to the discernible effect on $\sigma$, we note that providing an offset affects the normalized overlap minimally over the frequency range $95\leqslant N \leqslant 130\ \text{\text{rpm}}$ and $\lambda$ remains  between $11-24\%$. 

\section{Results: Simulations}
\label{sec:results-S}
We now relate the experimental results obtained above to the interfacial pressure-gradient mechanism of Bhateja et al.\cite{bhateja2017}  For this we will perform simulations based on the discrete element method\cite{cundall79} (DEM). The computational technique, the procedure followed, and chosen parameters are stated in Appendix~\ref{App:A}. We record  that  the size ratio of grains is kept at $1.4$. We recall from  Sec.~\ref{sec:intro} that as per the interfacial pressure-gradient mechanism of Bhateja et al. \cite{bhateja2017} the interfacial pressure between the top and the bottom layers develops a gradient that  drives the bottom layer from the right to the left end of the channel, thereby segregating the  mixture. In our experiments the large and small grains occupied, respectively, the top and bottom layers after the initial vertical sorting and, so, separate to the right and left ends of the channel. 

\begin{figure}[b!]
\centering
\includegraphics[scale=0.071]{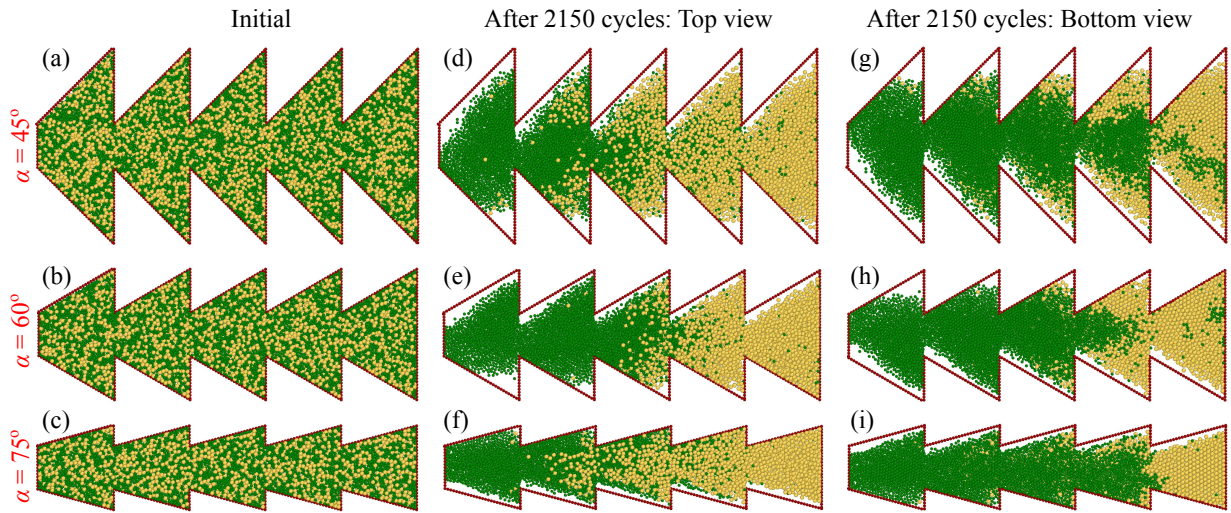}
\caption{Initial and segregated configurations of binary granular mixtures shaken at 105 \text{rpm} at three taper angles. Top and bottom views are shown, with big and small grains depicted in yellow and green, respectively. Snapshots are captured while the channel is in motion.}
\label{fig:snapsTop}
\end{figure}

We simulate segregation of binary granular mixtures in channels having taper angles $\alpha=45^\circ, 60^\circ$ and $75^\circ$. Simulation snapshots for these taper angles depicting the initial and segregated configurations in steady state after 2150 cycles are displayed in Fig.~\ref{fig:snapsTop}. We note qualitative similarity with experimental observations of the previous section. Moreover, comparing the views of the top and the bottom of the channel in Fig.~\ref{fig:snapsTop} confirms the variation in the segregation through the depth of the mixture. This supports our computing an overlap length in experiments. 

\begin{figure}[b!]
\centering
\includegraphics[scale=0.25]{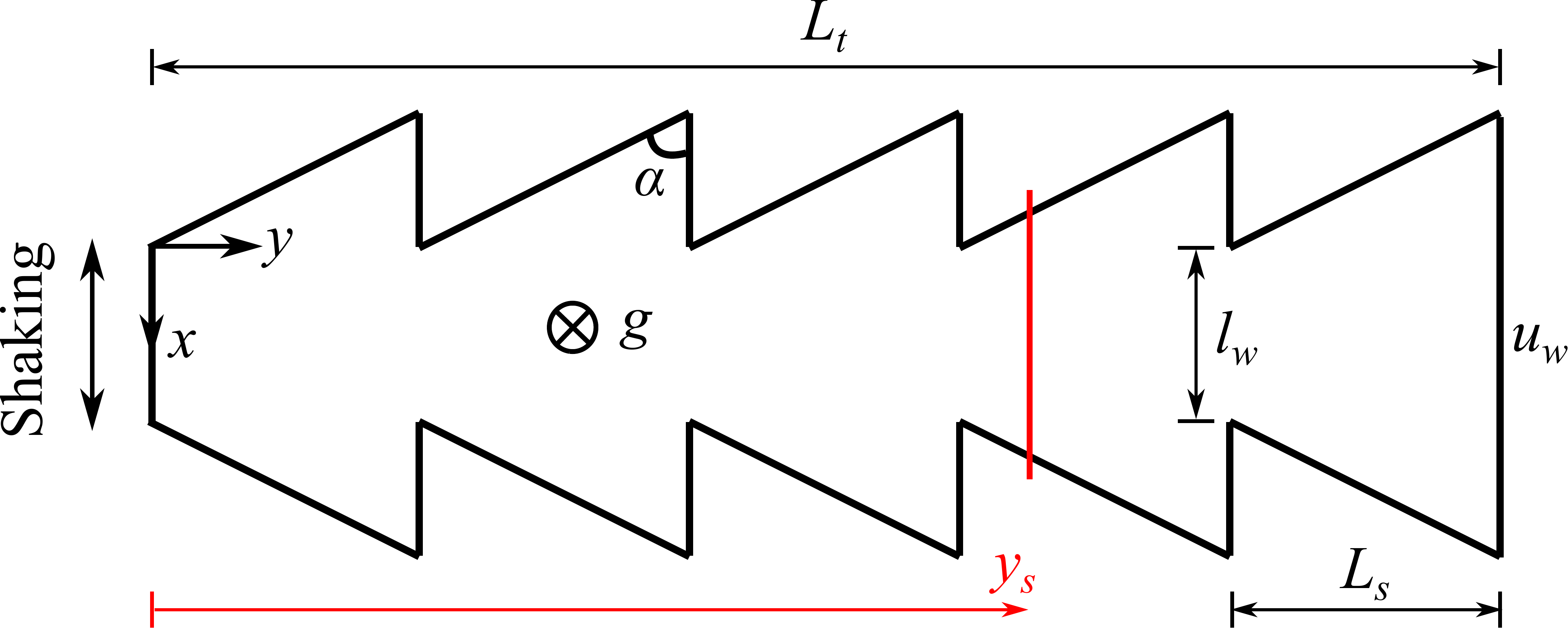}
\caption{A schematic showing the lower and upper widths of a trapezium, and the axial location $y_s$ beyond which a certain percentage $\psi$ of small grains lie. Refer to the main text and Appendix~A for details. The $z$-axis is normal to the channel base and is directed out of the plane of paper. Note that we choose different length variables for simulations in order to avoid any confusion with experiments; cf.~Fig.~\ref{fig:setup}.}
\label{fig:simSetup}
\end{figure}

\begin{figure*}[t!]
\centering
\includegraphics[width=\textwidth]{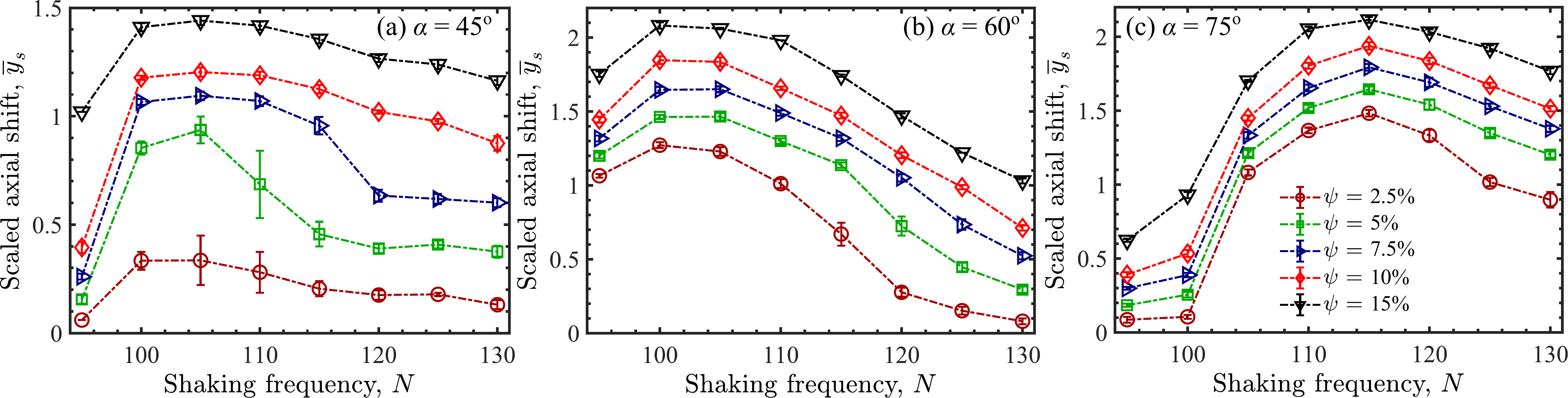}
\caption{Variation of the scaled axial shift $\overline{y}_s$ with $N$ for several values of $\psi$ for (a) $\alpha=45^\circ$, (b) $\alpha=60^\circ$, and (c) $\alpha=75^\circ$. The scaled axial shift is averaged over 51 shaking cycles, ranging from 2100 to 2150 cycles. Legend for all plots is given in (c).}
\label{fig:sas}
\end{figure*}

We will quantify segregation in our simulations through two measures. We first define a {\em bulk segregation index} $\sigma_s$  by
\begin{equation}
\label{eq:sigmas}
\sigma_s = 2(y_b^c - y_s^c)/L_t,
\end{equation}
where $y_b^c$ and $y_s^c$ are axial locations of the centers of mass of big and small grains, respectively.  A larger $\sigma_s$ will generally indicate a better separated granular aggregate. However, $\sigma_s$ does not capture the spread of a species about its mass center; e.g.  it is possible that $\sigma_s$ remain high but a significant fraction of small grains continue to occupy the rightmost trapezoidal section, and we will present an example at the end of this section illustrating this. Thus, we introduce a second segregation measure. For this we first define the percentage ratio $\psi=100\times (N_s/n_s)$ of the number of small grains $N_s$ located to the right of an axial location $y_s$, with respect to the total number of small grains $n_s$ in the mixture; Fig.~\ref{fig:simSetup} shows $y_s$ in a sketch. Thus, fixing a percentage ratio $\psi$ yields a {\em scaled axial shift}
\begin{equation}
	\overline{y}_s = (L_t-y_s)/L_s,
\end{equation}	
where $L_s$ is the length of a trapezium (see Fig.~\ref{fig:simSetup}); this is the second measure of segregation that we will employ. For any $\psi$, we note that $\overline{y}_s=r$ implies that less than $\psi$ percent of small grains lie in the rightmost $r$ trapeziums.  Thus, we may say that a mixture is {\em not} well separated if ${\overline y}_s < 1$. Figure~\ref{fig:sas} presents the variation of $\overline{y}_s$ with $N$ for three taper angles. We observe that $\overline{y}_s$ increases monotonically with $\psi$ for a fixed shaking frequency. Furthermore, as shown in Fig.~\ref{fig:sas}, qualitatively similar results are obtained for $\psi$ varying between $2.5\%$ and $15\%$. Therefore,  without loss of generality, we will set $\psi=5\%$, and assume that, if the corresponding $\overline{y}_s \geqslant 1$, then we have achieved an acceptable level of axial separation. Below we discuss further  how $\sigma_s$ and ${\overline y}_s$ relate to each other and with the segregation measure $\sigma$ utilized in experiments.

\begin{figure}[!ht]
\centering
\includegraphics[scale=0.6]{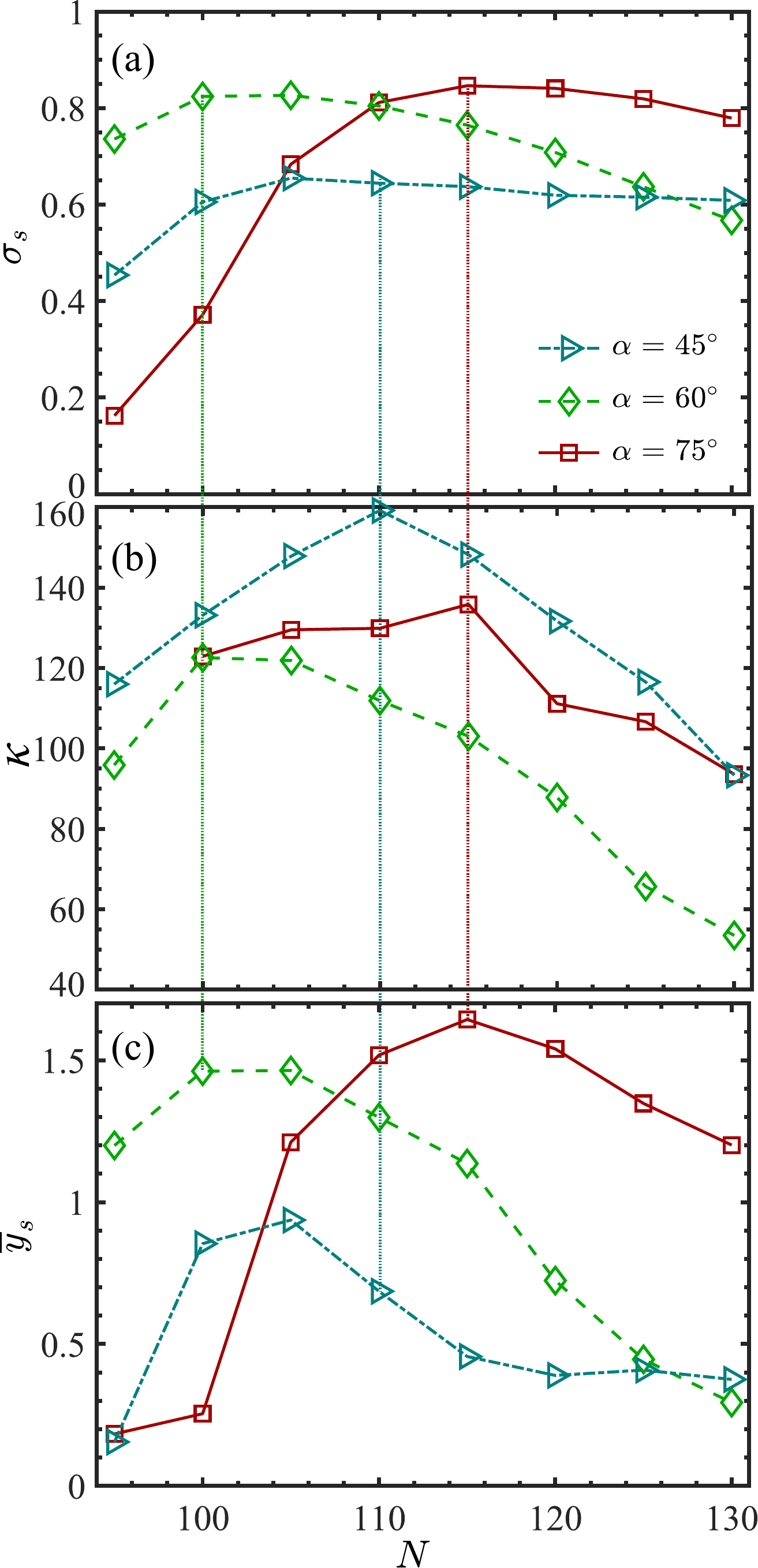}
\caption{Variation with shaking frequency $N$ of the (a) bulk segregation index $\sigma_s$, (b) inverse $\kappa$ of the scaled pressure gradient  and (c) scaled axial shift $\overline{y}_s$. Channels with three different taper angles $\alpha$ are investigated.  Vertical lines passing through the peak of $\kappa$ are drawn to indicate correlations with $\sigma_s$ and ${\overline y}_s$. Legend for all plots is given in (a).}
\label{fig:SI_pgrad}
\end{figure}

The variation of $\sigma_s$ with shaking frequency $N$ for $\alpha=45^\circ, 60^\circ$ and $75^\circ$ is shown in Fig.~\ref{fig:SI_pgrad}(a).  Here, $\sigma_s$ is calculated utilizing the axial center of mass locations averaged over sufficient number of cycles. As in experiments, see, e.g. Fig.~\ref{fig:MM_alpha_SigLam}a, the bulk segregation index shows a non-monotonic trend with shaking frequency for these three angles. It grows with the shaking frequency, attains a maximum value and reduces on further increase in $N$.

%

Consider next the variation of {\em scaled pressure gradient} ${\overline{P'}} = P'/\rho A \omega^2$, where $P'=dP/dy$ is the appropriately non-dimensionalized pressure gradient at the interface of the top and bottom layer of grains that forms after the rapid vertical sorting; recall Fig.~\ref{fig:cartoon} and the associated discussion. The procedure for computing pressure gradient is outlined in Appendix~\ref{App:press}. The variation of $\kappa = 1/\overline{P'}$ with $N$ is shown in Fig.~\ref{fig:SI_pgrad}(b) for various taper angles $\alpha$. Note that plotting the inverse of the scaled pressure gradient does not affect our analysis in any way, and is done so as to limit the vertical span of the plots. For each $\alpha$, it is clear that the behaviour of $\kappa$ is non-monotonic with $N$, similarly to what is exhibited by $\sigma_s$ and the experimental segregation index $\sigma$. The simulation with $N=95$ \text{rpm} in the $75^\circ$ channel should not be taken into consideration given the absence of significant granular flow in the system. In this case the bottom layer comprising small grains does not reverse its motion but, rather, remains roughly static, while the top layer consisting of big grains displaces by five-grain diameters towards the right. For this reason the pressure gradient for this case is not plotted. Returning to Fig.~\ref{fig:SI_pgrad}, guided by the vertical lines through the peaks of $\kappa$, we observe that the frequency response of  $\sigma_s$  matches well with that of $\kappa$, thereby demonstrating a correlation between the bulk segregation index and the scaled pressure gradient.  

Finally,  in Fig.~\ref{fig:SI_pgrad}(c) we plot the second segregation measure, the scaled axial shift $\overline{y}_s$, with shaking frequency $N$. We find that the curve for $\overline{y}_s$ correlates very well with the scaled pressure gradient through its inverse $\kappa$ and the bulk segregation index $\sigma_s$. Indeed, the largest values of $\sigma_s$ and $\overline{y}_s$  for $\alpha=75^\circ$ occur at the same frequency. The correspondence between the two segregation measures and the scaled pressure gradient  helps lend further support to the interfacial pressure-gradient mechanism of Bhateja et al.\citep{bhateja2017}.

%
%
\begin{figure}[b!]
\centering
\includegraphics[scale=0.6]{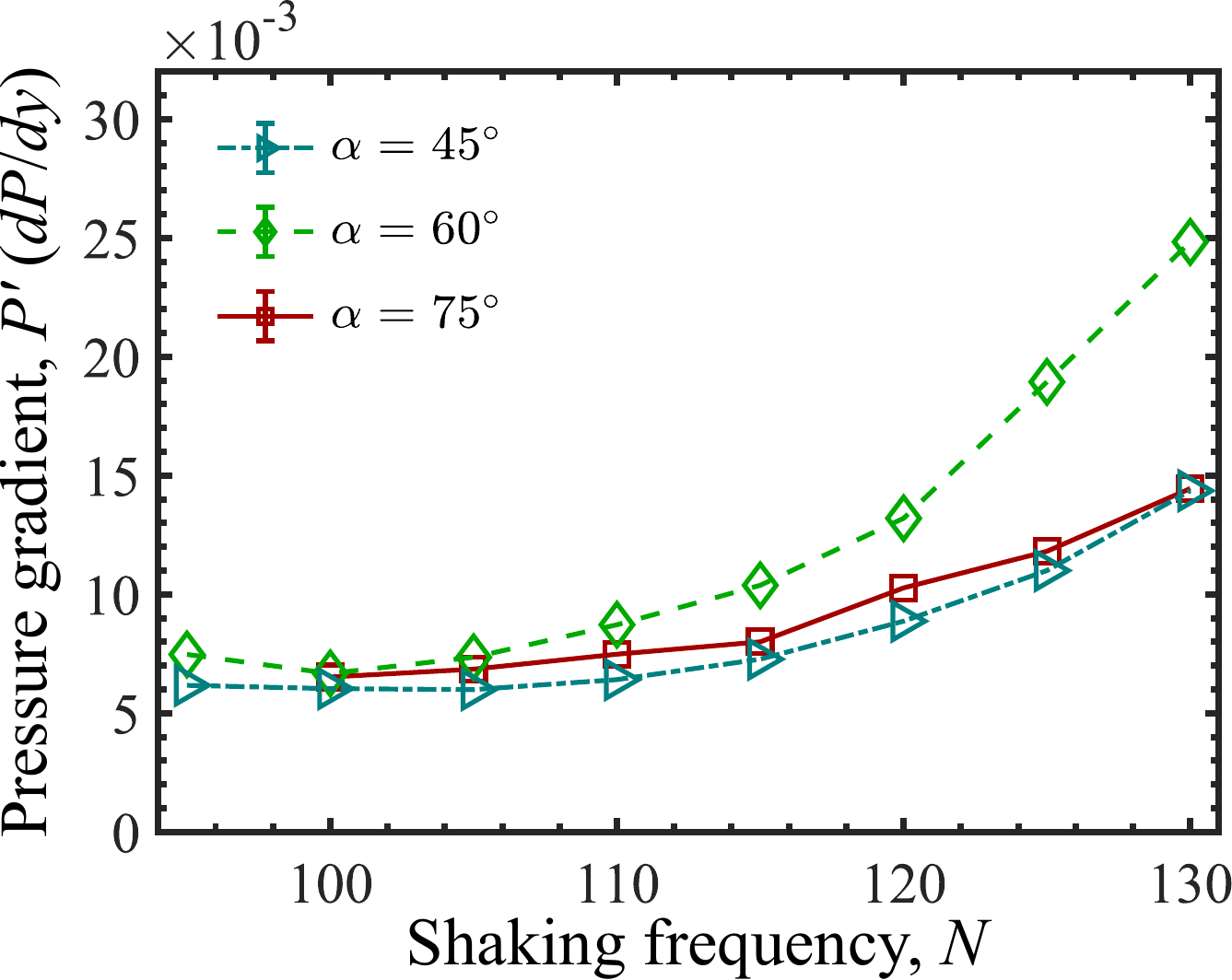}
\caption{Variation of the non-dimensionalized pressure gradient $P'=dP/dy$ with shaking frequency $N$. Channels with three different taper angles $\alpha$ are investigated.}
\label{fig:pgrad}
\end{figure}

Before closing this section we make three remarks. First, we explain why the scaled pressure gradient is the correct physical parameter to correlate with the segregation process rather than the pressure gradient alone. This is because the segregation is driven by an interplay between the interfacial pressure gradient that drives the grains at the bottom leftwards and the rightward motion provided by the tapered side walls, the extent of which is proportional to $A\omega^2$.  Indeed, consider  the variation of pressure gradient $P'$ with shaking frequency $N$ presented in Fig.~\ref{fig:pgrad}. We note that, for all taper angles the pressure gradient rises with shaking frequencies whereas, in contrast, the segregation index $\sigma_s$ reduces with increasing $N$ beyond an optimal frequency -- compare, say, Figs.~\ref{fig:SI_pgrad}a and \ref{fig:pgrad}. At the same time, the inverse $\kappa$ of the scaled pressure gradient ${\overline P}'$ in Fig.~\ref{fig:SI_pgrad}b reaches a maximum at about the same $N$ as $\sigma_s$ and then reduces. This correlation between $\kappa$ and $\sigma_s$ corresponds to our expectation that at low shaking frequencies the sidewalls are unable to drive enough of the grains towards the channel's right end. This prevents the formation of an adequate pressure gradient that would drive the bottom layer of grains to the left. On the other hand, when $N$ is high enough, then the rightward push from the walls would dominate and all grains would be pushed in that direction, suppressing axial segregation and, thereby, lowering $\sigma_s$.
%

\begin{figure}[!t]
\centering
\includegraphics[scale=0.6]{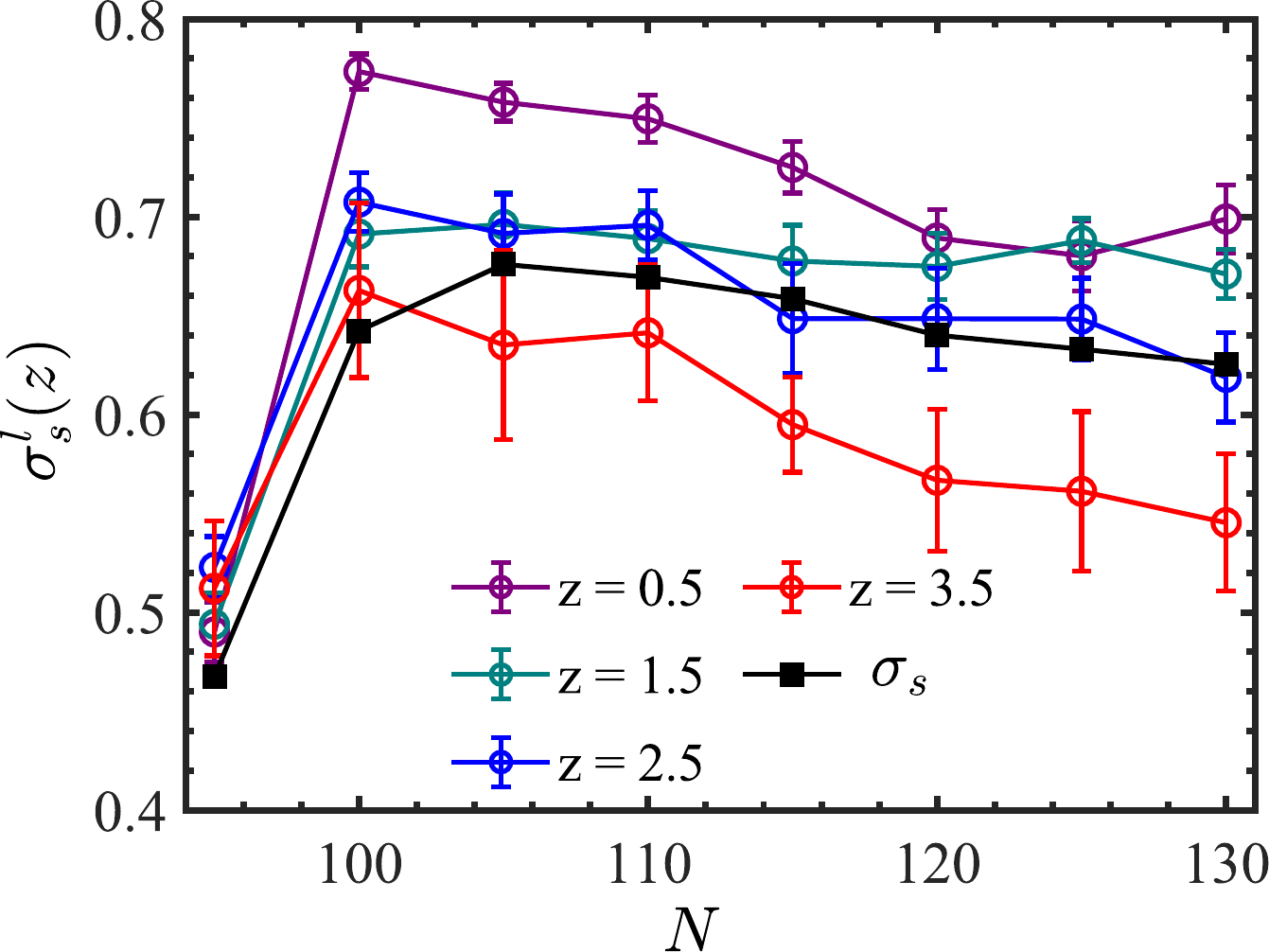}
\caption{Layerwise segregation index $\sigma^l_s(z)$   at different depths $z$ (scaled by $d_s$) versus shaking frequency $N$ in a $\alpha = 45^\circ$ channel. The initial fill height is nearly 3.6.}
\label{fig:SI_z_45}
\end{figure}

Second, we note that the bulk segregation index $\sigma_s$ relates to the entire collection of grains, as it employs the locations of the centers of mass of the small and big grains in the system. In contrast, the experimentally determined segregation index $\sigma$ considers grains in the top layer alone. In order to check how $\sigma$ and $\sigma_s$  relate to each other, we estimated a {\em layerwise}  segregation index $\sigma^l_s(z)$ that computes the separation between the mass centers of the big and small grains in a bin that is  centered at the vertical location $z$, has a vertical thickness of one particle diameter $d_s$ and spans the length and width of the channel. Thus, $\sigma^l_s(z)$ may be computed through a formula analogous to \eqref{eq:sigmas}, with $\sigma^l_s$ for the topmost bin corresponding to $\sigma$. Figure~\ref{fig:SI_z_45}  plots the steady state value of $\sigma^l_s(z)$ at various $z$ coordinates with $N$  in a channel with $\alpha=45^\circ$. We find that the variation of $\sigma^l_s(z)$ is qualitatively similar at all depths so that $\sigma_s$, which is the through-depth average of $\sigma^l_s(z)$,  will also behave analogously. Therefore, generally, $\sigma_s$ found from simulations may be employed as a surrogate for $\sigma$ estimated from experiments.

%

\begin{figure}[b!]
\centering
\includegraphics[scale=0.125]{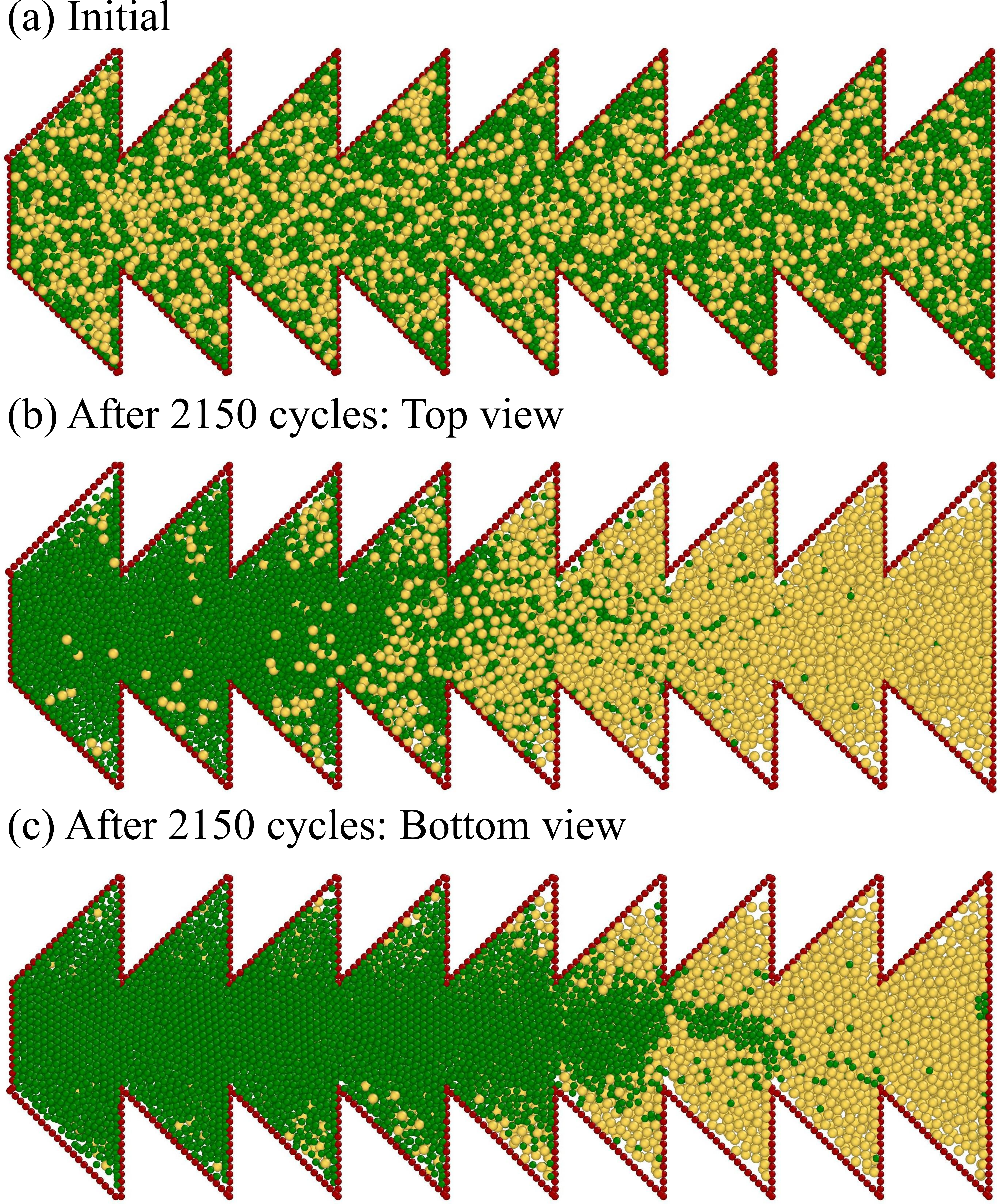}
\caption{Snapshots displaying the initial and segregated configurations of a binary granular mixture in an $\alpha = 45^\circ$ channel having nine trapezoidal sections. The mixture is shaken at 105 \text{rpm}. Big and small grains are displayed in yellow and green, respectively.}
\label{fig:snaps45SE}
\end{figure}
%

\begin{figure}[t!]
\centering
\includegraphics[scale=0.6]{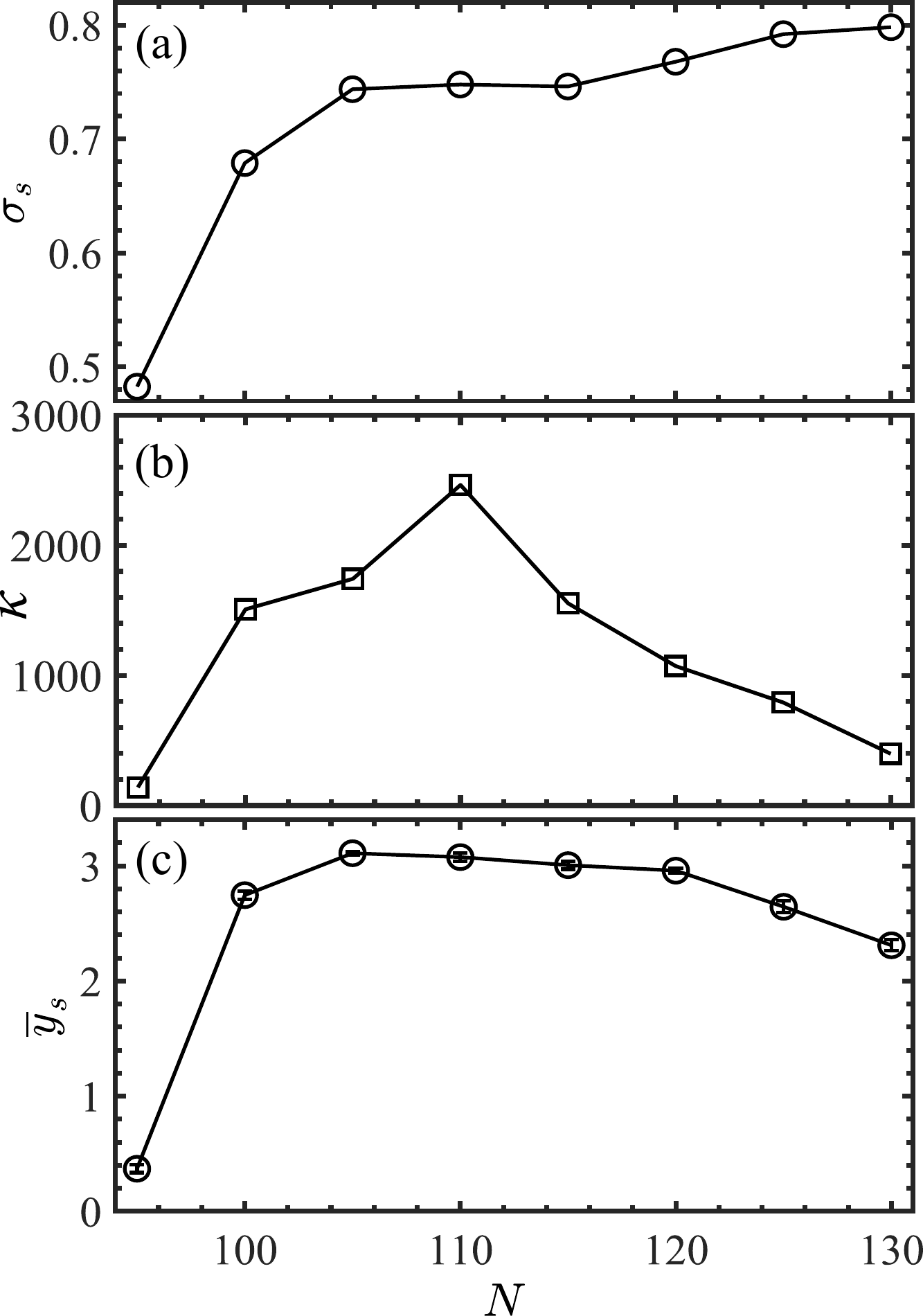}
\caption{Variation with shaking frequency $N$ of the (a) bulk segregation index $\sigma_s$, (b) inverse of the scaled pressure gradient $\kappa$, and (c) scaled axial shift $\overline{y}_s$  in the $\alpha = 45^\circ$ channel shown in Fig.~\ref{fig:snaps45SE}.}
\label{fig:taper45SE}
\end{figure}

Finally, we comment on how the channel geometry is changed when the taper angle $\alpha$ is varied and its effect on the segregation process. In the simulations reported above, while varying $\alpha$, the number of sections, the channel length $L_t$, and the length $l_w$ of the shorter of the parallel sides of the trapezoidal sections remain invariant; see Fig.~\ref{fig:simSetup} and associated discussion in Appendix~\ref{App:A}. Accordingly, the length $u_w$  of the longer parallel side of the trapeziums changes. In contrast, recall, in experiments the number of sections are varied as $\alpha$ changes, while $l_e, u_e$ and $L$ are kept the same. We now investigate the effect that this difference may have. For this we  simulate segregation in a $45^\circ$ channel with nine trapezoidal sections, similarly to the experiments shown in  Fig.~\ref{fig:MM45_75}.  Simulation snapshots for segregation in the nine-section channel are depicted in Fig.~\ref{fig:snaps45SE}, and  Fig.~\ref{fig:taper45SE} presents the corresponding variations in $\sigma_s$, $\kappa$ and $\overline{y}_s$ with $N$. The bulk segregation index $\sigma_s$ reaches a high value by $N = 105$ rpm, beyond which it stays flat before increasing slightly yet again when $N\geqslant120$ rpm. At the same time, the scaled shift ${\overline y}_s$  reaches a peak of about 3 by $N = 105$ rpm after which it decays slightly. Recall that ${\overline y}_s = 3$ indicates that 5\% of the small grains are in the rightmost three trapezoidal sections.  The inverse of the scaled pressure gradient $\kappa$ correlates reasonably with both segregation measures, but cannot explain the second increase in $\sigma_s$ at $N=120$ rpm. We believe that this is because, as discussed at the beginning of this section, the bulk segregation index $\sigma_s$ is unable to adequately capture the distribution of a species about its mass center, which is estimated better by the scaled shift ${\overline y}_s$. Indeed, a decrease in ${\overline y}_s$ suggests a lowering in the segregation quality which, in Fig.~\ref{fig:taper45SE}(c), happens markedly after $N \geqslant 120$ rpm. In contrast, $\sigma_s$ in Fig.~\ref{fig:taper45SE}(a) grows when $N \geqslant 120$ rpm, falsely indicating improved segregation.


\section{Conclusions}
\label{sec:conclude}
Axial segregation in binary mixtures in a laterally shaken multi-trapezium channel was originally reported by Bhateja et al.\citep{bhateja2017}. In this paper, we carried that investigation further in two important ways. First, we studied experimentally the influence on the segregation process of various system parameters, namely, the size ratio of the grains, the shaking frequency and the channel's geometry through the taper and the offset of the sidewalls. To quantify the segregation found experimentally, we introduced two measures: the segregation index $\sigma$ that quantified the separation between big and small grains from the top view of the system, and the overlap length $\lambda$ that reported the through-depth variation in the segregation quality. We found that segregation quality depends on the shaking frequency $N$, and there are generally different, but closely spaced, optimal frequencies at which $\sigma$ and $\lambda$ are extremized. However, because the variation of $\lambda$ is gentle around its optimal $N$, so one may locate a unique optimal $N$ that provides the best segregation. This  optimal frequency is observed to lower with increase in size ratio. Further, this feature is true for all except the very low taper angles $\alpha$. When $\alpha = 30^\circ$ segregation is weak because of the formation of large interacting dead zones in the narrow trapezoidal sections. Anomalously, at $\alpha = 15^\circ$ the segregation, instead of becoming poorer, reverses in direction and improves dramatically. This is an unexplained feature that requires further investigation. We also found that offsetting the sidewalls axially releases the dead zones and improves segregation quality. Lastly, we did not observe any dependence of the segregation quality on the shape of the particles, which corroborates a recent finding by Jones et al. \citep{jones2020}, wherein they showed that the degree of segregation in a binary mixture of different shaped particles depends mainly on the volume ratio of the two species, irrespective of their shapes.

In the second part of the work, we simulated the segregation process in order to relate the experimental observations to the interfacial pressure gradient mechanism of Bhateja et al.\citep{bhateja2017}. For this we utilized discrete element simulations. The segregation quality was again quantified through two measures better suited to simulations: a bulk segregation index $\sigma_s$ that reported the separation between the centers of mass of the small and big grains, and a scaled shift ${\overline y}_s$ that estimated the location of a segregation boundary in the axial direction. We showed that both $\sigma_s$ and ${\overline y}_s$ generally correspond with each other. Further,  as far as qualitative response is concerned, $\sigma_s$ may be utilized as a surrogate for the experimental segregation measure $\sigma$. Finally, we demonstrated that the segregation quality  correlated very well with the scaled interfacial pressure gradient ${\overline {P'}}$  -- the ratio of the interfacial pressure gradient to the axial body force provided by the sidewalls --  in that the best segregation was achieved  when ${\overline {P'}}$ reached an extremum. This provided further support to the interfacial pressure gradient mechanism.

Looking ahead there are several avenues to investigate. For example, it appears possible to utilize depth-averaged granular flow equations to theoretically model the segregation process. Furthermore, Bhateja et al.\citep{bhateja2017} reported that the present system can segregate mixtures with more than two species, and this needs to be studied more carefully.

%
\appendix
\section{Computational setup and procedure}
\label{App:A}
Computations are performed utilizing an in-house code based on the discrete element method\cite{cundall79,shafer1996}. Grains are assumed to be cohesionless and dry spheres. Interactions between grains occur when they come into contact, and the contact force is computed by considering the linear spring-dashpot model\cite{zhang1996,bkm2003}. For modelling the normal contact force, both spring and dashpot are considered, whereas only dashpot is taken into account for the tangential interaction. Details of the model are given elsewhere\cite{bhateja2016}. Note that we do not consider the attractive force during rebounding phase\cite{poschel2005}. All quantities of interest presented in this work are dimensionless, which are nondimensionalized in terms of the grain diameter $d_s$, mass density $\rho_s$, and gravitational acceleration $g$.

The shaking amplitude is taken to be the same as in experiments, i.e. $A=7$ cm. The simulations are performed for shaking frequencies ranging between 95 \text{rpm} and 130 \text{rpm}, in steps of 5 \text{rpm}. The dimensionless shaking acceleration $\Gamma$ varies between 0.71 and 1.32 for these amplitude and frequency range. Recall, as argued in Sec.~\ref{sec:exp-a}, we present the data in terms of $N$ as the shaking amplitude remains constant. The size ratio of grains is 1.4, with the diameter of small ones being $d_s=5$ mm, which corresponds to the binary mixture of peas and green gram (see Tables~\ref{tab:table1} and \ref{tab:table2}). All grains have the same density. 

All simulations are performed in a regular multi-trapezium channel; see Fig.~\ref{fig:simSetup}. The number of trapeziums in the channel is 5, unless explicitly stated otherwise. Three taper angles are considered, $\alpha=45^\circ, 60^\circ,$ and $75^\circ$.  For a channel   with $\alpha=60^\circ$ the parameters are as follows. The lower and upper straight edges of a trapezium, in terms of $d_s$, are $l_w=14d_s$ and $u_w=43d_s$, respectively. In dimensional units these correspond to $l_w=70$ mm and $u_w=215$ mm, which are smaller than those employed in experiments. Nevertheless, $l_w$ is large enough to ensure an uninterrupted flow\cite{zuriguel2005} through the passage at the junction of two trapeziums. The number of small and big grains are $n_s=5000$ and $n_b=1825$, respectively. 

While changing the taper angle, we ensure that the length of each section, the lower width, and the initial filling fraction of the mixture remain the same. Therefore, the width of the upper edge $(u_w)$ and the number of small $(n_s)$ and big $(n_b)$ grains are varied accordingly for $\alpha=45^\circ$ and $75^\circ$. Employing the format [$u_w, n_s, n_b$]  these are,  respectively, [64.23, 6865, 2505] and [27.46, 3640, 1330]. The initial fill height of the mixtures for all cases is $3.6 \pm 0.1$.

The normal spring constant is $k_n=10^6$, and the restitution and friction coefficients for grain-grain interactions are 0.8 and 0.3, respectively. The same values of these parameters are employed for wall-grain interactions. Note that the normal damping coefficient $c_n$ may be computed from the restitution coefficient $e_n$ and the effective mass $m_e = m_i m_j/{m_i + m_j}$ of two colliding grains $i$ and $j$ having mass $m_i$ and $m_j$, respectively:
\begin{equation}
c_n = \frac{2\, \sqrt{(\text{ln}\, e_n)^2 \, m_e\,k_n}}{(\pi^2+(\text{ln}\, e_n)^2)^{1/2}}.
\end{equation}
In this study, the normal and tangential damping coefficients are equal.

\section{Interfacial pressure gradient}
\label{App:press}

\begin{figure*}[t!]
\centering
\includegraphics[scale=0.15]{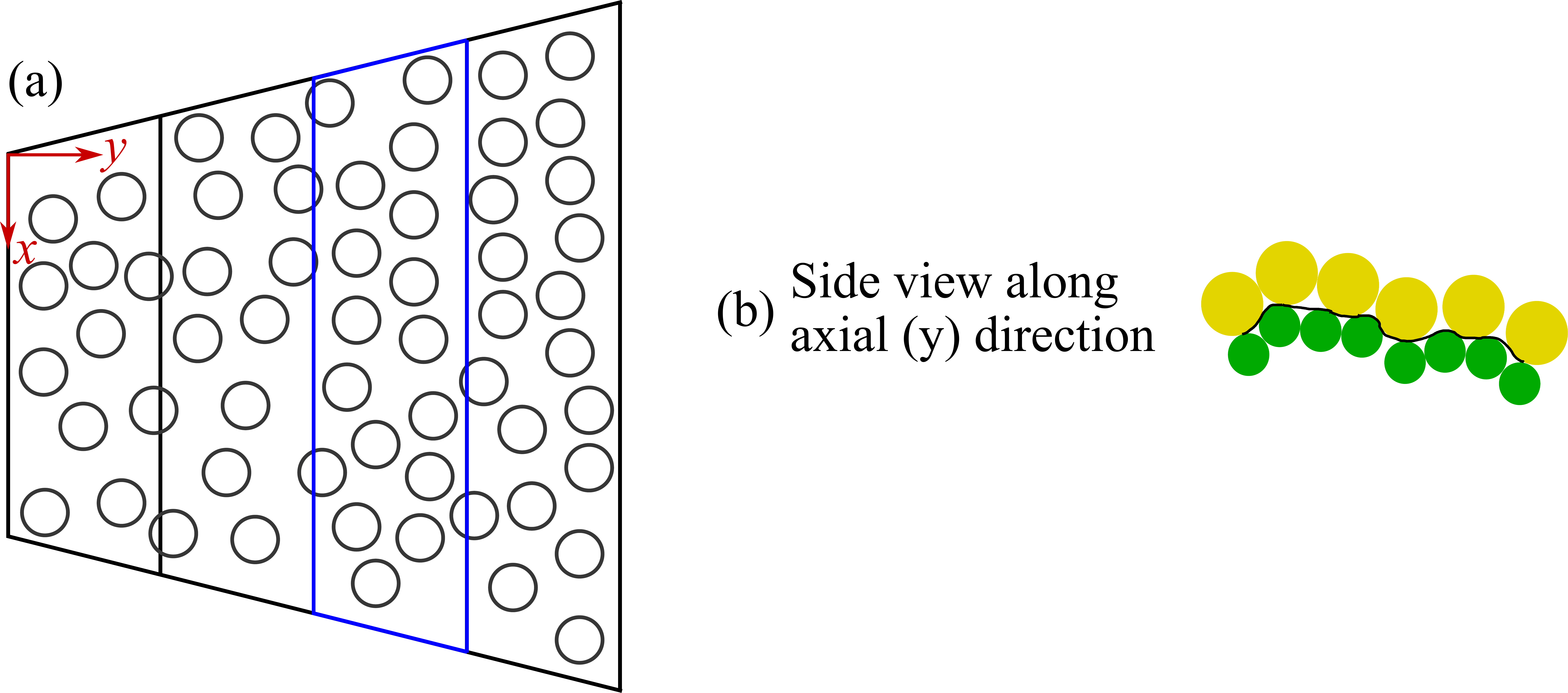}
\caption{(a) An illustration for the discretization of a trapezoidal section of the multi-trapezium channel along the axial ($y$) direction. The region outlined in blue represents a typical bin in which collisional and streaming stress between large and small grains is calculated. (b) A schematic showing interface between the contacting big and small grains.}
\label{fig:pgradCalc}
\end{figure*}

For computing pressure at the interface of top and bottom layers comprising big and small grains, respectively, the channel is first divided into several trapezoidal bins along axial direction as illustrated schematically in Fig.~\ref{fig:pgradCalc}(a). In each bin, contacts between those big and small grains are identified whenever the former is located above the latter as depicted in Fig.~\ref{fig:pgradCalc}(b). This helps in identifying the interface between the top and bottom layers that are formed after the rapid vertical sorting; cf. Fig.~\ref{fig:cartoon}. It is evident that the interface of the contacting grains will not be straight. However, as a first approximation, the interface area may be taken equal to the area $A_b$ of a trapezium, as shown in blue in Fig.~\ref{fig:pgradCalc}(a). The pressure due to collisions $P_c$ is calculated in each bin by obtaining the total vertical force (along $z$ direction) exerted by big grains on the small ones and dividing it by the interface area:
\begin{equation}
P_c = \frac{1}{A_b}\sum_{j=1}^{N_c} f_{j}^{z},
\end{equation}
where $N_c$ is the total number of contacts between the big and small grains, and $f_{j}^z$ is the vertical component of the force acting at the contact $j$.

In addition to the collisional component of the pressure, the streaming or kinetic contribution due to fluctuation velocity of grains in each bin is also estimated through the tensor\cite{campbell1989,tripathi2010}
\begin{equation}
\bm{\sigma}^k = \frac{1}{V_b} \sum_{i=1}^{N_b} m_i \bm{c}_i \otimes \bm{c}_i,
\end{equation}
where $N_b$ is the number of contacting big and small grains enclosed in a volume $V_b$ of height equal to one small particle diameter centered at the interface of the top and bottom layers, and $m_i$ and $\bm{c}_i=\bm{u}_i - \bm{v}$ are the mass and fluctuational velocity of grain $i$ with $\bm{u}_i$ being its instantaneous velocity and $\bm{v}$ the mean velocity of grains located in $V_b$. Note that the  non-dimensional height of the volume $V_b$ is one, so that  $V_b=A_b$.

\begin{figure}[t!]
\centering
\includegraphics[scale=0.65]{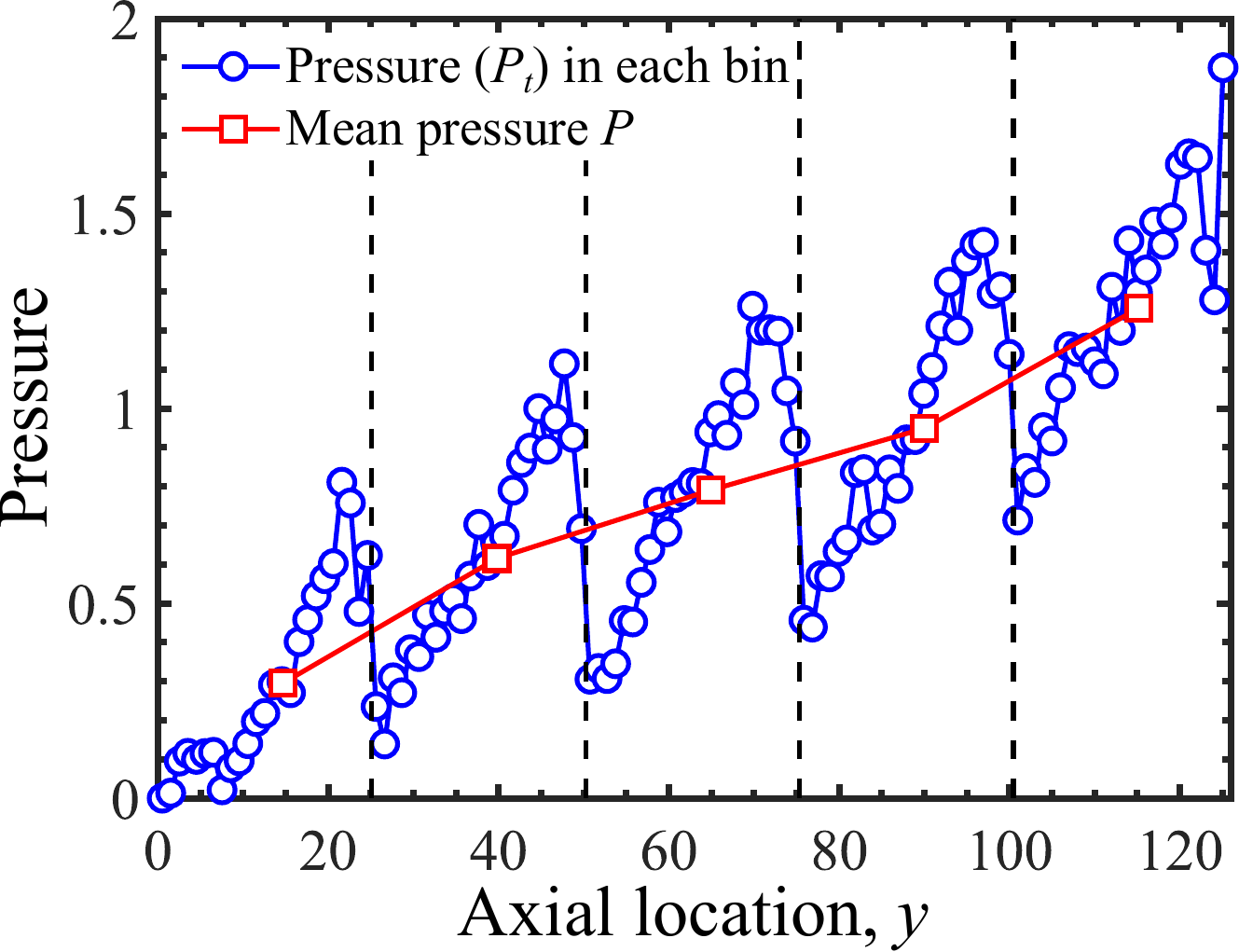}
\caption{Axial variation of pressure and its mean value in each trapezoidal section. The data are shown for $\alpha=60^\circ$ after 33 shaking cycles. The mixture is shaken at 105 rpm. Qualitatively similar variation is obtained for other taper angles. The dashed line represents the junction of two neighbouring trapezia.}
\label{fig:press}
\end{figure}

The total pressure in each trapezoidal bin is obtained by adding the $zz$-component of $\bm{\sigma}^{k}$ to the pressure due to collisions, i.e. $P_t = P_c + \sigma_{zz}^{k}$. It is worth mentioning that the kinetic component is small in comparison to $P_c$, which is expected for dense granular flows\cite{campbell1989}. Specifically, we note that the maximum value of $\sigma_{zz}^k$ is less than $1\%$ of $P_c$ for all shaking frequencies and taper angles considered. The variation of pressure along the axial direction is displayed in Fig.~\ref{fig:press} for the $60^\circ$ channel shaken at 105 rpm. We observe that the pressure $P_t$ grows locally in each trapezoidal section, besides its overall increase along the channel. The steep drop in the pressure at the right end of each section is caused due to the sudden reduction in the channel's width. From the pressure data we compute the mean pressure $P$ in each trapezoidal section. Subsequently, the pressure gradient $P' = dP/dy$ along the channel is estimated from a linear fit to these mean pressure values.

\begin{figure}[t!]
\centering
\includegraphics[scale=0.65]{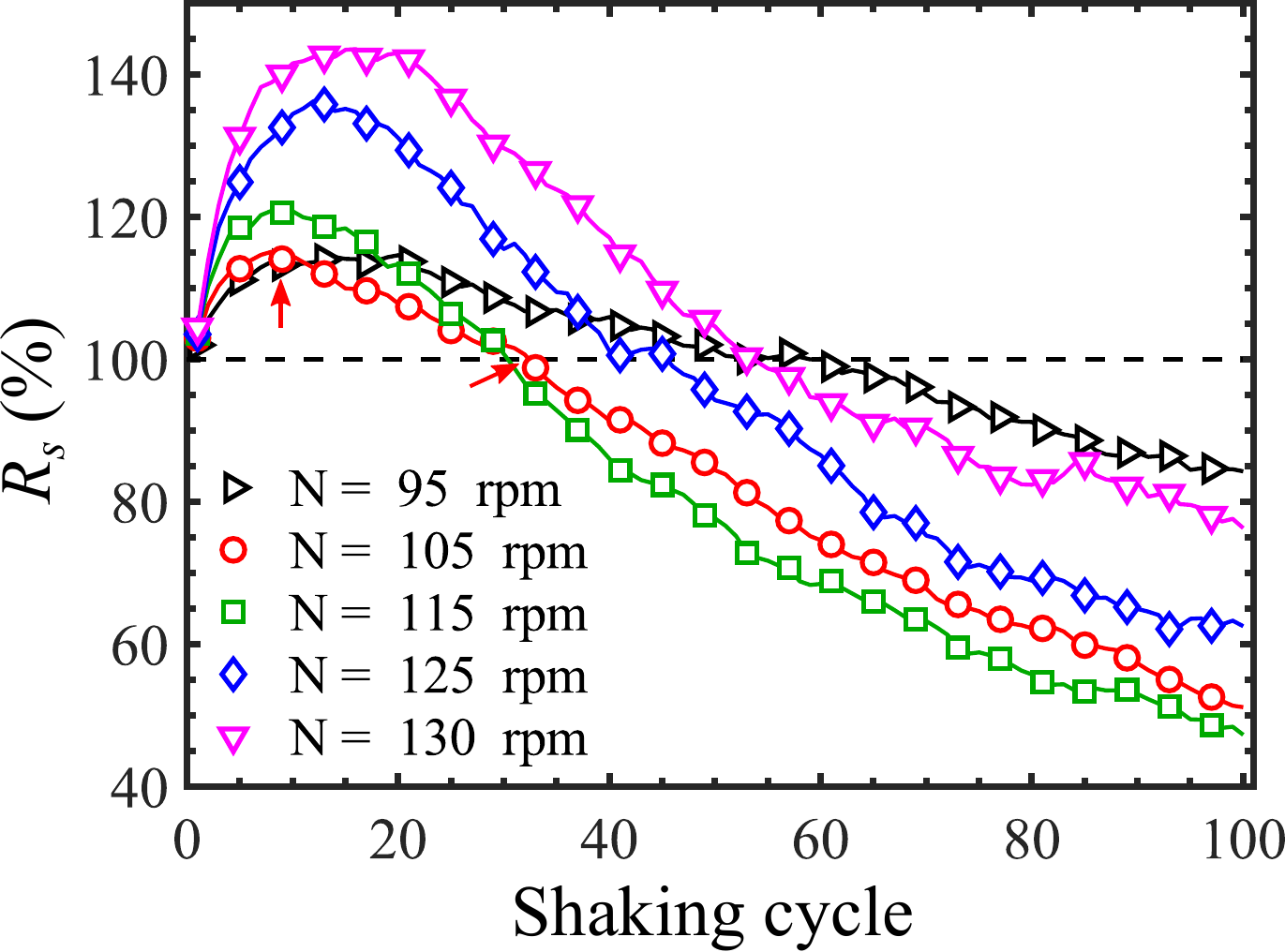}
\caption{Variation of the percentage fraction $R_s$ of small grains in the rightmost section with shaking cycle in a  $\alpha=60^\circ$ channel. Qualitatively similar variation is obtained for all other frequencies and channels. Red arrows indicate the shaking cycles corresponding to when $R_s$ is maximum $R_s$ and when returns to being   $100\%$ at the  shaking frequency $N=105$ rpm.}
\label{fig:numdens}
\end{figure}

The mean value of the pressure gradient  is itself computed by averaging $P'$ over several vibration cycles. To find this averaging range, we consider the percentage fraction $R_s = 100 \times (s_c/s_0)$ of small grains in the rightmost trapezoidal section, where $s_0$ and $s_c$ are the number of small grains present in the rightmost section before shaking and after $c$ vibration cycles, respectively. Thus, $R_s > 100\%$ indicates that the number of small grains in the rightmost section is greater that it was at the beginning of the simulation. The variation of $R_s$ with shaking cycle is shown in Fig.~\ref{fig:numdens} for various frequencies. In all cases, as time progresses $R_s$  first increases, attains a peak, and then decreases. The range of vibration cycles over which to calculate the average pressure gradient is taken to be the number of cycles lying it takes for $R_s$  to reduce from its peak to  $100\%$, i.e. when the number of small grains equals to its value at the start of the simulation. Besides 100\%, we also considered averages taken over a greater number of cycles corresponding to when $R_s$  returned to 95\% and 90\% from its peak value. The variation of the mean pressure gradient with shaking frequency $N$ was found to display the same trends.

\newpage
\section*{Acknowledgements}
We gratefully acknowledge the support of IIT Kanpur for providing access to its HPC facility. M.I.H.A. sincerely acknowledges the financial support from IIT Kanpur through post-doctoral fellowship. A.B. gratefully acknowledges the financial support from IIT Goa through Start-up Grant (2019/SG/AB/025). We also thank  Satya Prakash Mishra from IIT Kanpur for help with experiments.

\section*{Conflict of Interest}
The authors have no conflicts to disclose.
\section*{Data Availability}
The data that support the findings of this study are available from the corresponding author upon reasonable request.

\bibliography{ashish_bib.bib}

\providecommand{\noopsort}[1]{}\providecommand{\singleletter}[1]{#1}%
\begin{thebibliography}{39}%
\makeatletter
\providecommand \@ifxundefined [1]{%
 \@ifx{#1\undefined}
}%
\providecommand \@ifnum [1]{%
 \ifnum #1\expandafter \@firstoftwo
 \else \expandafter \@secondoftwo
 \fi
}%
\providecommand \@ifx [1]{%
 \ifx #1\expandafter \@firstoftwo
 \else \expandafter \@secondoftwo
 \fi
}%
\providecommand \natexlab [1]{#1}%
\providecommand \enquote  [1]{``#1''}%
\providecommand \bibnamefont  [1]{#1}%
\providecommand \bibfnamefont [1]{#1}%
\providecommand \citenamefont [1]{#1}%
\providecommand \href@noop [0]{\@secondoftwo}%
\providecommand \href [0]{\begingroup \@sanitize@url \@href}%
\providecommand \@href[1]{\@@startlink{#1}\@@href}%
\providecommand \@@href[1]{\endgroup#1\@@endlink}%
\providecommand \@sanitize@url [0]{\catcode `\\12\catcode `\$12\catcode
  `\&12\catcode `\#12\catcode `\^12\catcode `\_12\catcode `\%12\relax}%
\providecommand \@@startlink[1]{}%
\providecommand \@@endlink[0]{}%
\providecommand \url  [0]{\begingroup\@sanitize@url \@url }%
\providecommand \@url [1]{\endgroup\@href {#1}{\urlprefix }}%
\providecommand \urlprefix  [0]{URL }%
\providecommand \Eprint [0]{\href }%
\providecommand \doibase [0]{http://dx.doi.org/}%
\providecommand \selectlanguage [0]{\@gobble}%
\providecommand \bibinfo  [0]{\@secondoftwo}%
\providecommand \bibfield  [0]{\@secondoftwo}%
\providecommand \translation [1]{[#1]}%
\providecommand \BibitemOpen [0]{}%
\providecommand \bibitemStop [0]{}%
\providecommand \bibitemNoStop [0]{.\EOS\space}%
\providecommand \EOS [0]{\spacefactor3000\relax}%
\providecommand \BibitemShut  [1]{\csname bibitem#1\endcsname}%
\let\auto@bib@innerbib\@empty
\bibitem [{\citenamefont {Ottino}\ and\ \citenamefont
  {Khakhar}(2000)}]{ottino_khakhar2000}%
  \BibitemOpen
  \bibfield  {author} {\bibinfo {author} {\bibfnamefont {J.~M.}\ \bibnamefont
  {Ottino}}\ and\ \bibinfo {author} {\bibfnamefont {D.~V.}\ \bibnamefont
  {Khakhar}},\ }\bibfield  {title} {\enquote {\bibinfo {title} {Mixing and
  segregation of granular materials},}\ }\href@noop {} {\bibfield  {journal}
  {\bibinfo  {journal} {Annu. Rev. Fluid Mech.}\ }\textbf {\bibinfo {volume}
  {32}},\ \bibinfo {pages} {55--91} (\bibinfo {year} {2000})}\BibitemShut
  {NoStop}%
\bibitem [{\citenamefont {Kudrolli}(2004)}]{kudrolli2004}%
  \BibitemOpen
  \bibfield  {author} {\bibinfo {author} {\bibfnamefont {A.}~\bibnamefont
  {Kudrolli}},\ }\bibfield  {title} {\enquote {\bibinfo {title} {Size
  separation in vibrated granular matter},}\ }\href@noop {} {\bibfield
  {journal} {\bibinfo  {journal} {Rep. Prog. Phys.}\ }\textbf {\bibinfo
  {volume} {67}},\ \bibinfo {pages} {209} (\bibinfo {year} {2004})}\BibitemShut
  {NoStop}%
\bibitem [{\citenamefont {Gray}(2018)}]{gray2018}%
  \BibitemOpen
  \bibfield  {author} {\bibinfo {author} {\bibfnamefont {J.~M. N.~T.}\
  \bibnamefont {Gray}},\ }\bibfield  {title} {\enquote {\bibinfo {title}
  {Particle segregation in dense granular flows},}\ }\href@noop {} {\bibfield
  {journal} {\bibinfo  {journal} {Annu. Rev. Fluid Mech.}\ }\textbf {\bibinfo
  {volume} {50}},\ \bibinfo {pages} {407--433} (\bibinfo {year}
  {2018})}\BibitemShut {NoStop}%
\bibitem [{\citenamefont {Rosato}\ \emph {et~al.}(1987)\citenamefont {Rosato},
  \citenamefont {Strandburg}, \citenamefont {Prinz},\ and\ \citenamefont
  {Swendsen}}]{rosato87}%
  \BibitemOpen
  \bibfield  {author} {\bibinfo {author} {\bibfnamefont {A.}~\bibnamefont
  {Rosato}}, \bibinfo {author} {\bibfnamefont {K.~J.}\ \bibnamefont
  {Strandburg}}, \bibinfo {author} {\bibfnamefont {F.}~\bibnamefont {Prinz}}, \
  and\ \bibinfo {author} {\bibfnamefont {R.~H.}\ \bibnamefont {Swendsen}},\
  }\bibfield  {title} {\enquote {\bibinfo {title} {Why the brazil nuts are on
  top: Size segregation of particulate matter by shaking},}\ }\href@noop {}
  {\bibfield  {journal} {\bibinfo  {journal} {Phys. Rev. Lett.}\ }\textbf
  {\bibinfo {volume} {58}},\ \bibinfo {pages} {1038--1041} (\bibinfo {year}
  {1987})}\BibitemShut {NoStop}%
\bibitem [{\citenamefont {Shinbrot}\ and\ \citenamefont
  {Muzzio}(1998)}]{shinbrot1998}%
  \BibitemOpen
  \bibfield  {author} {\bibinfo {author} {\bibfnamefont {T.}~\bibnamefont
  {Shinbrot}}\ and\ \bibinfo {author} {\bibfnamefont {F.~J.}\ \bibnamefont
  {Muzzio}},\ }\bibfield  {title} {\enquote {\bibinfo {title} {Reverse buoyancy
  in shaken granular beds},}\ }\href@noop {} {\bibfield  {journal} {\bibinfo
  {journal} {Phys. Rev. Lett.}\ }\textbf {\bibinfo {volume} {81}},\ \bibinfo
  {pages} {4365--4368} (\bibinfo {year} {1998})}\BibitemShut {NoStop}%
\bibitem [{\citenamefont {Liao}\ \emph {et~al.}(2014)\citenamefont {Liao},
  \citenamefont {Hunt}, \citenamefont {Hsiau},\ and\ \citenamefont
  {Lu}}]{liao2014}%
  \BibitemOpen
  \bibfield  {author} {\bibinfo {author} {\bibfnamefont {C.~C.}\ \bibnamefont
  {Liao}}, \bibinfo {author} {\bibfnamefont {M.~L.}\ \bibnamefont {Hunt}},
  \bibinfo {author} {\bibfnamefont {S.~S.}\ \bibnamefont {Hsiau}}, \ and\
  \bibinfo {author} {\bibfnamefont {S.~H.}\ \bibnamefont {Lu}},\ }\bibfield
  {title} {\enquote {\bibinfo {title} {Investigation of the effect of a bumpy
  base on granular segregation and transport properties under vertical
  vibration},}\ }\href@noop {} {\bibfield  {journal} {\bibinfo  {journal}
  {Phys. Fluids}\ }\textbf {\bibinfo {volume} {26}},\ \bibinfo {pages} {073302}
  (\bibinfo {year} {2014})}\BibitemShut {NoStop}%
\bibitem [{\citenamefont {Schnautz}\ \emph {et~al.}(2005)\citenamefont
  {Schnautz}, \citenamefont {Brito}, \citenamefont {Kruelle},\ and\
  \citenamefont {Rehberg}}]{schnautz2005}%
  \BibitemOpen
  \bibfield  {author} {\bibinfo {author} {\bibfnamefont {T.}~\bibnamefont
  {Schnautz}}, \bibinfo {author} {\bibfnamefont {R.}~\bibnamefont {Brito}},
  \bibinfo {author} {\bibfnamefont {C.~A.}\ \bibnamefont {Kruelle}}, \ and\
  \bibinfo {author} {\bibfnamefont {I.}~\bibnamefont {Rehberg}},\ }\bibfield
  {title} {\enquote {\bibinfo {title} {A horizontal brazil-nut effect and its
  reverse},}\ }\href@noop {} {\bibfield  {journal} {\bibinfo  {journal} {Phys.
  Rev. Lett.}\ }\textbf {\bibinfo {volume} {95}},\ \bibinfo {pages} {028001}
  (\bibinfo {year} {2005})}\BibitemShut {NoStop}%
\bibitem [{\citenamefont {Mobarakabadi}\ \emph {et~al.}(2013)\citenamefont
  {Mobarakabadi}, \citenamefont {Oskoee}, \citenamefont {Schr{\"o}ter},\ and\
  \citenamefont {Habibi}}]{mobarakabadi2013}%
  \BibitemOpen
  \bibfield  {author} {\bibinfo {author} {\bibfnamefont {S.}~\bibnamefont
  {Mobarakabadi}}, \bibinfo {author} {\bibfnamefont {E.~N.}\ \bibnamefont
  {Oskoee}}, \bibinfo {author} {\bibfnamefont {M.}~\bibnamefont
  {Schr{\"o}ter}}, \ and\ \bibinfo {author} {\bibfnamefont {M.}~\bibnamefont
  {Habibi}},\ }\bibfield  {title} {\enquote {\bibinfo {title} {Granular
  transport in a horizontally vibrated sawtooth channel},}\ }\href@noop {}
  {\bibfield  {journal} {\bibinfo  {journal} {Phys. Rev. E}\ }\textbf {\bibinfo
  {volume} {88}},\ \bibinfo {pages} {042201} (\bibinfo {year}
  {2013})}\BibitemShut {NoStop}%
\bibitem [{\citenamefont {Bhateja}, \citenamefont {Sharma},\ and\ \citenamefont
  {Singh}(2017)}]{bhateja2017}%
  \BibitemOpen
  \bibfield  {author} {\bibinfo {author} {\bibfnamefont {A.}~\bibnamefont
  {Bhateja}}, \bibinfo {author} {\bibfnamefont {I.}~\bibnamefont {Sharma}}, \
  and\ \bibinfo {author} {\bibfnamefont {J.~K.}\ \bibnamefont {Singh}},\
  }\bibfield  {title} {\enquote {\bibinfo {title} {Segregation physics of a
  macroscale granular ratchet},}\ }\href@noop {} {\bibfield  {journal}
  {\bibinfo  {journal} {Phys. Rev. Fluids}\ }\textbf {\bibinfo {volume} {2}},\
  \bibinfo {pages} {052301} (\bibinfo {year} {2017})}\BibitemShut {NoStop}%
\bibitem [{\citenamefont {Mobarakabadi}\ \emph {et~al.}(2017)\citenamefont
  {Mobarakabadi}, \citenamefont {Adrang}, \citenamefont {Habibi},\ and\
  \citenamefont {Oskoee}}]{mobarakabadi2017}%
  \BibitemOpen
  \bibfield  {author} {\bibinfo {author} {\bibfnamefont {S.}~\bibnamefont
  {Mobarakabadi}}, \bibinfo {author} {\bibfnamefont {N.}~\bibnamefont
  {Adrang}}, \bibinfo {author} {\bibfnamefont {M.}~\bibnamefont {Habibi}}, \
  and\ \bibinfo {author} {\bibfnamefont {E.~N.}\ \bibnamefont {Oskoee}},\
  }\bibfield  {title} {\enquote {\bibinfo {title} {Segregation of a binary
  granular mixture in a vibrating sawtooth base container},}\ }\href@noop {}
  {\bibfield  {journal} {\bibinfo  {journal} {Eur. Phys. J. E}\ }\textbf
  {\bibinfo {volume} {40}},\ \bibinfo {pages} {1--7} (\bibinfo {year}
  {2017})}\BibitemShut {NoStop}%
\bibitem [{\citenamefont {Richard}\ and\ \citenamefont
  {Taberlet}(2008)}]{richard2008}%
  \BibitemOpen
  \bibfield  {author} {\bibinfo {author} {\bibfnamefont {P.}~\bibnamefont
  {Richard}}\ and\ \bibinfo {author} {\bibfnamefont {N.}~\bibnamefont
  {Taberlet}},\ }\bibfield  {title} {\enquote {\bibinfo {title} {Recent
  advances in dem simulations of grains in a rotating drum},}\ }\href@noop {}
  {\bibfield  {journal} {\bibinfo  {journal} {Soft Matter}\ }\textbf {\bibinfo
  {volume} {4}},\ \bibinfo {pages} {1345--1348} (\bibinfo {year}
  {2008})}\BibitemShut {NoStop}%
\bibitem [{\citenamefont {Yang}\ \emph
  {et~al.}(2017{\natexlab{a}})\citenamefont {Yang}, \citenamefont {Sun},
  \citenamefont {Zhang},\ and\ \citenamefont {Chew}}]{yang2017a}%
  \BibitemOpen
  \bibfield  {author} {\bibinfo {author} {\bibfnamefont {S.}~\bibnamefont
  {Yang}}, \bibinfo {author} {\bibfnamefont {Y.}~\bibnamefont {Sun}}, \bibinfo
  {author} {\bibfnamefont {L.}~\bibnamefont {Zhang}}, \ and\ \bibinfo {author}
  {\bibfnamefont {J.~W.}\ \bibnamefont {Chew}},\ }\bibfield  {title} {\enquote
  {\bibinfo {title} {Numerical study on the axial segregation dynamics of a
  binary-size granular mixture in a three-dimensional rotating drum},}\
  }\href@noop {} {\bibfield  {journal} {\bibinfo  {journal} {Phys. Fluids}\
  }\textbf {\bibinfo {volume} {29}},\ \bibinfo {pages} {103302} (\bibinfo
  {year} {2017}{\natexlab{a}})}\BibitemShut {NoStop}%
\bibitem [{\citenamefont {Yang}\ \emph
  {et~al.}(2017{\natexlab{b}})\citenamefont {Yang}, \citenamefont {Zhang},
  \citenamefont {Luo},\ and\ \citenamefont {Chew}}]{yang2017b}%
  \BibitemOpen
  \bibfield  {author} {\bibinfo {author} {\bibfnamefont {S.}~\bibnamefont
  {Yang}}, \bibinfo {author} {\bibfnamefont {L.}~\bibnamefont {Zhang}},
  \bibinfo {author} {\bibfnamefont {K.}~\bibnamefont {Luo}}, \ and\ \bibinfo
  {author} {\bibfnamefont {J.~W.}\ \bibnamefont {Chew}},\ }\bibfield  {title}
  {\enquote {\bibinfo {title} {Dem study of the size-induced segregation
  dynamics of a ternary-size granular mixture in the rolling-regime rotating
  drum},}\ }\href@noop {} {\bibfield  {journal} {\bibinfo  {journal} {Phys.
  Fluids}\ }\textbf {\bibinfo {volume} {29}},\ \bibinfo {pages} {123301}
  (\bibinfo {year} {2017}{\natexlab{b}})}\BibitemShut {NoStop}%
\bibitem [{\citenamefont {Staron}\ and\ \citenamefont
  {Phillips}(2014)}]{staron2014}%
  \BibitemOpen
  \bibfield  {author} {\bibinfo {author} {\bibfnamefont {L.}~\bibnamefont
  {Staron}}\ and\ \bibinfo {author} {\bibfnamefont {J.~C.}\ \bibnamefont
  {Phillips}},\ }\bibfield  {title} {\enquote {\bibinfo {title} {Segregation
  time-scale in bi-disperse granular flows},}\ }\href@noop {} {\bibfield
  {journal} {\bibinfo  {journal} {Phys. Fluids}\ }\textbf {\bibinfo {volume}
  {26}},\ \bibinfo {pages} {033302} (\bibinfo {year} {2014})}\BibitemShut
  {NoStop}%
\bibitem [{\citenamefont {Mandal}\ and\ \citenamefont
  {Khakhar}(2019)}]{mandal2019}%
  \BibitemOpen
  \bibfield  {author} {\bibinfo {author} {\bibfnamefont {S.}~\bibnamefont
  {Mandal}}\ and\ \bibinfo {author} {\bibfnamefont {D.~V.}\ \bibnamefont
  {Khakhar}},\ }\bibfield  {title} {\enquote {\bibinfo {title} {Dense granular
  flow of mixtures of spheres and dumbbells down a rough inclined plane:
  Segregation and rheology},}\ }\href@noop {} {\bibfield  {journal} {\bibinfo
  {journal} {Phys. Fluids}\ }\textbf {\bibinfo {volume} {31}},\ \bibinfo
  {pages} {023304} (\bibinfo {year} {2019})}\BibitemShut {NoStop}%
\bibitem [{\citenamefont {Tripathi}\ and\ \citenamefont
  {Khakhar}(2013)}]{tripathi2013}%
  \BibitemOpen
  \bibfield  {author} {\bibinfo {author} {\bibfnamefont {A.}~\bibnamefont
  {Tripathi}}\ and\ \bibinfo {author} {\bibfnamefont {D.~V.}\ \bibnamefont
  {Khakhar}},\ }\bibfield  {title} {\enquote {\bibinfo {title} {Density
  difference-driven segregation in a dense granular flow},}\ }\href@noop {}
  {\bibfield  {journal} {\bibinfo  {journal} {J. Fluid Mech.}\ }\textbf
  {\bibinfo {volume} {717}},\ \bibinfo {pages} {643--669} (\bibinfo {year}
  {2013})}\BibitemShut {NoStop}%
\bibitem [{\citenamefont {Yan}\ \emph {et~al.}(2003)\citenamefont {Yan},
  \citenamefont {Shi}, \citenamefont {Hou}, \citenamefont {Lu},\ and\
  \citenamefont {Chan}}]{yan2003}%
  \BibitemOpen
  \bibfield  {author} {\bibinfo {author} {\bibfnamefont {X.}~\bibnamefont
  {Yan}}, \bibinfo {author} {\bibfnamefont {Q.}~\bibnamefont {Shi}}, \bibinfo
  {author} {\bibfnamefont {M.}~\bibnamefont {Hou}}, \bibinfo {author}
  {\bibfnamefont {K.}~\bibnamefont {Lu}}, \ and\ \bibinfo {author}
  {\bibfnamefont {C.~K.}\ \bibnamefont {Chan}},\ }\bibfield  {title} {\enquote
  {\bibinfo {title} {Effects of air on the segregation of particles in a shaken
  granular bed},}\ }\href@noop {} {\bibfield  {journal} {\bibinfo  {journal}
  {Phys. Rev. Lett.}\ }\textbf {\bibinfo {volume} {91}},\ \bibinfo {pages}
  {014302} (\bibinfo {year} {2003})}\BibitemShut {NoStop}%
\bibitem [{\citenamefont {Grossman}(1997)}]{grossman1997}%
  \BibitemOpen
  \bibfield  {author} {\bibinfo {author} {\bibfnamefont {E.~L.}\ \bibnamefont
  {Grossman}},\ }\bibfield  {title} {\enquote {\bibinfo {title} {Effects of
  container geometry on granular convection},}\ }\href@noop {} {\bibfield
  {journal} {\bibinfo  {journal} {Phys. Rev. E.}\ }\textbf {\bibinfo {volume}
  {56}},\ \bibinfo {pages} {3290--3300} (\bibinfo {year} {1997})}\BibitemShut
  {NoStop}%
\bibitem [{\citenamefont {Savage}\ and\ \citenamefont {Lun}(1988)}]{savage88}%
  \BibitemOpen
  \bibfield  {author} {\bibinfo {author} {\bibfnamefont {S.~B.}\ \bibnamefont
  {Savage}}\ and\ \bibinfo {author} {\bibfnamefont {C.~K.~K.}\ \bibnamefont
  {Lun}},\ }\bibfield  {title} {\enquote {\bibinfo {title} {Particle size
  segregation in inclined chute flow of dry cohesionless granular solids},}\
  }\href@noop {} {\bibfield  {journal} {\bibinfo  {journal} {J. Fluid Mech.}\
  }\textbf {\bibinfo {volume} {189}},\ \bibinfo {pages} {311--335} (\bibinfo
  {year} {1988})}\BibitemShut {NoStop}%
\bibitem [{\citenamefont {Knight}, \citenamefont {Jaeger},\ and\ \citenamefont
  {Nagel}(1993)}]{knight1993}%
  \BibitemOpen
  \bibfield  {author} {\bibinfo {author} {\bibfnamefont {J.~B.}\ \bibnamefont
  {Knight}}, \bibinfo {author} {\bibfnamefont {H.~M.}\ \bibnamefont {Jaeger}},
  \ and\ \bibinfo {author} {\bibfnamefont {S.~R.}\ \bibnamefont {Nagel}},\
  }\bibfield  {title} {\enquote {\bibinfo {title} {Vibration induced size
  separation in granular media: {T}he convection connection},}\ }\href@noop {}
  {\bibfield  {journal} {\bibinfo  {journal} {Phys. Rev. Lett.}\ }\textbf
  {\bibinfo {volume} {70}},\ \bibinfo {pages} {3728--3731} (\bibinfo {year}
  {1993})}\BibitemShut {NoStop}%
\bibitem [{\citenamefont {Fan}\ and\ \citenamefont {Hill}(2011)}]{fan2011}%
  \BibitemOpen
  \bibfield  {author} {\bibinfo {author} {\bibfnamefont {Y.}~\bibnamefont
  {Fan}}\ and\ \bibinfo {author} {\bibfnamefont {K.~M.}\ \bibnamefont {Hill}},\
  }\bibfield  {title} {\enquote {\bibinfo {title} {Phase transitions in
  shear-induced segregation of granular materials},}\ }\href@noop {} {\bibfield
   {journal} {\bibinfo  {journal} {Phys. Rev. Lett.}\ }\textbf {\bibinfo
  {volume} {106}},\ \bibinfo {pages} {218301} (\bibinfo {year}
  {2011})}\BibitemShut {NoStop}%
\bibitem [{\citenamefont {Umbanhowar}, \citenamefont {Lueptow},\ and\
  \citenamefont {Ottino}(2019)}]{umbanhowar2019}%
  \BibitemOpen
  \bibfield  {author} {\bibinfo {author} {\bibfnamefont {P.~B.}\ \bibnamefont
  {Umbanhowar}}, \bibinfo {author} {\bibfnamefont {R.~M.}\ \bibnamefont
  {Lueptow}}, \ and\ \bibinfo {author} {\bibfnamefont {J.~M.}\ \bibnamefont
  {Ottino}},\ }\bibfield  {title} {\enquote {\bibinfo {title} {Modeling
  segregation in granular flows},}\ }\href@noop {} {\bibfield  {journal}
  {\bibinfo  {journal} {Annu. Rev. Chem. Biomol. Eng}\ }\textbf {\bibinfo
  {volume} {10}},\ \bibinfo {pages} {129--153} (\bibinfo {year}
  {2019})}\BibitemShut {NoStop}%
\bibitem [{\citenamefont {M{\"o}bius}\ \emph {et~al.}(2001)\citenamefont
  {M{\"o}bius}, \citenamefont {Lauderdale}, \citenamefont {Nagel},\ and\
  \citenamefont {Jaeger}}]{mobius2001}%
  \BibitemOpen
  \bibfield  {author} {\bibinfo {author} {\bibfnamefont {M.~E.}\ \bibnamefont
  {M{\"o}bius}}, \bibinfo {author} {\bibfnamefont {B.~E.}\ \bibnamefont
  {Lauderdale}}, \bibinfo {author} {\bibfnamefont {S.~R.}\ \bibnamefont
  {Nagel}}, \ and\ \bibinfo {author} {\bibfnamefont {H.~M.}\ \bibnamefont
  {Jaeger}},\ }\bibfield  {title} {\enquote {\bibinfo {title} {Brazil-nut
  effect: {S}ize separation of granular particles},}\ }\href@noop {} {\bibfield
   {journal} {\bibinfo  {journal} {Nature}\ }\textbf {\bibinfo {volume}
  {414}},\ \bibinfo {pages} {270} (\bibinfo {year} {2001})}\BibitemShut
  {NoStop}%
\bibitem [{\citenamefont {Hong}, \citenamefont {Quinn},\ and\ \citenamefont
  {Luding}(2001)}]{hong2001}%
  \BibitemOpen
  \bibfield  {author} {\bibinfo {author} {\bibfnamefont {D.~C.}\ \bibnamefont
  {Hong}}, \bibinfo {author} {\bibfnamefont {P.~V.}\ \bibnamefont {Quinn}}, \
  and\ \bibinfo {author} {\bibfnamefont {S.}~\bibnamefont {Luding}},\
  }\bibfield  {title} {\enquote {\bibinfo {title} {Reverse {B}razil-nut
  problem: {C}ompetetion between percolation and condensation},}\ }\href@noop
  {} {\bibfield  {journal} {\bibinfo  {journal} {Phys. Rev. Lett.}\ }\textbf
  {\bibinfo {volume} {86}},\ \bibinfo {pages} {3423--3426} (\bibinfo {year}
  {2001})}\BibitemShut {NoStop}%
\bibitem [{\citenamefont {Shinbrot}(2004)}]{shinbrot2004}%
  \BibitemOpen
  \bibfield  {author} {\bibinfo {author} {\bibfnamefont {T.}~\bibnamefont
  {Shinbrot}},\ }\bibfield  {title} {\enquote {\bibinfo {title} {Granular
  materials: {T}he brazil nut effect -- in reverse},}\ }\href@noop {}
  {\bibfield  {journal} {\bibinfo  {journal} {Nature}\ }\textbf {\bibinfo
  {volume} {429}},\ \bibinfo {pages} {352--353} (\bibinfo {year}
  {2004})}\BibitemShut {NoStop}%
\bibitem [{mpd()}]{mpd}%
  \BibitemOpen
  \href@noop {} {\enquote {\bibinfo {title} {Machine and process design},}\
  }\bibinfo {howpublished} {\url{http://www.mpd-inc.com}},\ \bibinfo {note}
  {accessed: 08.03.2022}\BibitemShut {NoStop}%
\bibitem [{etb()}]{etb}%
  \BibitemOpen
  \href@noop {} {\enquote {\bibinfo {title} {Engineering tool box},}\ }\bibinfo
  {howpublished} {\url{https://www.engineeringtoolbox.com}},\ \bibinfo {note}
  {accessed: 08.03.2022}\BibitemShut {NoStop}%
\bibitem [{\citenamefont {Ansari}, \citenamefont {Rivas},\ and\ \citenamefont
  {Alam}(2018)}]{ansari2018}%
  \BibitemOpen
  \bibfield  {author} {\bibinfo {author} {\bibfnamefont {I.~H.}\ \bibnamefont
  {Ansari}}, \bibinfo {author} {\bibfnamefont {N.}~\bibnamefont {Rivas}}, \
  and\ \bibinfo {author} {\bibfnamefont {M.}~\bibnamefont {Alam}},\ }\bibfield
  {title} {\enquote {\bibinfo {title} {Phase-coexisting patterns, horizontal
  segregation, and controlled convection in vertically vibrated binary granular
  mixtures},}\ }\href {\doibase 10.1103/PhysRevE.97.012911} {\bibfield
  {journal} {\bibinfo  {journal} {Phys. Rev. E}\ }\textbf {\bibinfo {volume}
  {97}},\ \bibinfo {pages} {012911} (\bibinfo {year} {2018})}\BibitemShut
  {NoStop}%
\bibitem [{\citenamefont {Amon}\ \emph {et~al.}(2017)\citenamefont {Amon},
  \citenamefont {Born}, \citenamefont {Daniels}, \citenamefont {Dijksman},
  \citenamefont {Huang}, \citenamefont {Parker}, \citenamefont {Schr{\"o}ter},
  \citenamefont {Stannarius},\ and\ \citenamefont {Wierschem}}]{amon2017}%
  \BibitemOpen
  \bibfield  {author} {\bibinfo {author} {\bibfnamefont {A.}~\bibnamefont
  {Amon}}, \bibinfo {author} {\bibfnamefont {P.}~\bibnamefont {Born}}, \bibinfo
  {author} {\bibfnamefont {K.~E.}\ \bibnamefont {Daniels}}, \bibinfo {author}
  {\bibfnamefont {J.~A.}\ \bibnamefont {Dijksman}}, \bibinfo {author}
  {\bibfnamefont {K.}~\bibnamefont {Huang}}, \bibinfo {author} {\bibfnamefont
  {D.}~\bibnamefont {Parker}}, \bibinfo {author} {\bibfnamefont
  {M.}~\bibnamefont {Schr{\"o}ter}}, \bibinfo {author} {\bibfnamefont
  {R.}~\bibnamefont {Stannarius}}, \ and\ \bibinfo {author} {\bibfnamefont
  {A.}~\bibnamefont {Wierschem}},\ }\bibfield  {title} {\enquote {\bibinfo
  {title} {Preface: Focus on imaging methods in granular physics},}\
  }\href@noop {} {\bibfield  {journal} {\bibinfo  {journal} {Review of
  Scientific Instruments}\ }\textbf {\bibinfo {volume} {88}},\ \bibinfo {pages}
  {051701} (\bibinfo {year} {2017})}\BibitemShut {NoStop}%
\bibitem [{\citenamefont {Cundall}\ and\ \citenamefont
  {Strack}(1979)}]{cundall79}%
  \BibitemOpen
  \bibfield  {author} {\bibinfo {author} {\bibfnamefont {P.~A.}\ \bibnamefont
  {Cundall}}\ and\ \bibinfo {author} {\bibfnamefont {O.~D.~L.}\ \bibnamefont
  {Strack}},\ }\bibfield  {title} {\enquote {\bibinfo {title} {A discrete
  numerical model for granular assemblies},}\ }\href@noop {} {\bibfield
  {journal} {\bibinfo  {journal} {Geotechnique}\ }\textbf {\bibinfo {volume}
  {29(1)}},\ \bibinfo {pages} {47--65} (\bibinfo {year} {1979})}\BibitemShut
  {NoStop}%
\bibitem [{\citenamefont {Jones}\ \emph {et~al.}(2020)\citenamefont {Jones},
  \citenamefont {Ottino}, \citenamefont {Umbanhowar},\ and\ \citenamefont
  {Lueptow}}]{jones2020}%
  \BibitemOpen
  \bibfield  {author} {\bibinfo {author} {\bibfnamefont {R.~P.}\ \bibnamefont
  {Jones}}, \bibinfo {author} {\bibfnamefont {J.~M.}\ \bibnamefont {Ottino}},
  \bibinfo {author} {\bibfnamefont {P.~B.}\ \bibnamefont {Umbanhowar}}, \ and\
  \bibinfo {author} {\bibfnamefont {R.~M.}\ \bibnamefont {Lueptow}},\
  }\bibfield  {title} {\enquote {\bibinfo {title} {Remarkable simplicity in the
  prediction of nonspherical particle segregation},}\ }\href {\doibase
  10.1103/PhysRevResearch.2.042021} {\bibfield  {journal} {\bibinfo  {journal}
  {Phys. Rev. Research}\ }\textbf {\bibinfo {volume} {2}},\ \bibinfo {pages}
  {042021} (\bibinfo {year} {2020})}\BibitemShut {NoStop}%
\bibitem [{\citenamefont {Sh{\"a}fer}, \citenamefont {Dippel},\ and\
  \citenamefont {Wolf}(1996)}]{shafer1996}%
  \BibitemOpen
  \bibfield  {author} {\bibinfo {author} {\bibfnamefont {J.}~\bibnamefont
  {Sh{\"a}fer}}, \bibinfo {author} {\bibfnamefont {S.}~\bibnamefont {Dippel}},
  \ and\ \bibinfo {author} {\bibfnamefont {D.~E.}\ \bibnamefont {Wolf}},\
  }\bibfield  {title} {\enquote {\bibinfo {title} {Force schemes in simulations
  of granular materials},}\ }\href@noop {} {\bibfield  {journal} {\bibinfo
  {journal} {Journal de Physique I}\ }\textbf {\bibinfo {volume} {6}},\
  \bibinfo {pages} {5--20} (\bibinfo {year} {1996})}\BibitemShut {NoStop}%
\bibitem [{\citenamefont {Zhang}\ and\ \citenamefont
  {Whiten}(1996)}]{zhang1996}%
  \BibitemOpen
  \bibfield  {author} {\bibinfo {author} {\bibfnamefont {D.}~\bibnamefont
  {Zhang}}\ and\ \bibinfo {author} {\bibfnamefont {W.~J.}\ \bibnamefont
  {Whiten}},\ }\bibfield  {title} {\enquote {\bibinfo {title} {The calculation
  of contact forces between particles using spring and damping models},}\
  }\href@noop {} {\bibfield  {journal} {\bibinfo  {journal} {Powder Technol.}\
  }\textbf {\bibinfo {volume} {88}},\ \bibinfo {pages} {59--64} (\bibinfo
  {year} {1996})}\BibitemShut {NoStop}%
\bibitem [{\citenamefont {Mishra}(2003)}]{bkm2003}%
  \BibitemOpen
  \bibfield  {author} {\bibinfo {author} {\bibfnamefont {B.~K.}\ \bibnamefont
  {Mishra}},\ }\bibfield  {title} {\enquote {\bibinfo {title} {A review of
  computer simulation of tumbling mills by the discrete element method: Part
  {I}-contact mechanics},}\ }\href@noop {} {\bibfield  {journal} {\bibinfo
  {journal} {Int. J. Miner. Process.}\ }\textbf {\bibinfo {volume} {71}},\
  \bibinfo {pages} {73--93} (\bibinfo {year} {2003})}\BibitemShut {NoStop}%
\bibitem [{\citenamefont {Bhateja}, \citenamefont {Sharma},\ and\ \citenamefont
  {Singh}(2016)}]{bhateja2016}%
  \BibitemOpen
  \bibfield  {author} {\bibinfo {author} {\bibfnamefont {A.}~\bibnamefont
  {Bhateja}}, \bibinfo {author} {\bibfnamefont {I.}~\bibnamefont {Sharma}}, \
  and\ \bibinfo {author} {\bibfnamefont {J.~K.}\ \bibnamefont {Singh}},\
  }\bibfield  {title} {\enquote {\bibinfo {title} {Scaling of granular
  temperature in vibro-fluidized grains},}\ }\href@noop {} {\bibfield
  {journal} {\bibinfo  {journal} {Phys. Fluids}\ }\textbf {\bibinfo {volume}
  {28}},\ \bibinfo {pages} {043301} (\bibinfo {year} {2016})}\BibitemShut
  {NoStop}%
\bibitem [{\citenamefont {P{\"o}schel}\ and\ \citenamefont
  {Schwager}(2005)}]{poschel2005}%
  \BibitemOpen
  \bibfield  {author} {\bibinfo {author} {\bibfnamefont {T.}~\bibnamefont
  {P{\"o}schel}}\ and\ \bibinfo {author} {\bibfnamefont {T.}~\bibnamefont
  {Schwager}},\ }\href@noop {} {\emph {\bibinfo {title} {Computational Granular
  Dynamics: {M}odels and {A}lgorithms}}}\ (\bibinfo  {publisher} {Springer},\
  \bibinfo {year} {2005})\BibitemShut {NoStop}%
\bibitem [{\citenamefont {Zuriguel}\ \emph {et~al.}(2005)\citenamefont
  {Zuriguel}, \citenamefont {Garcimart{\'\i}n}, \citenamefont {Maza},
  \citenamefont {Pugnaloni},\ and\ \citenamefont {Pastor}}]{zuriguel2005}%
  \BibitemOpen
  \bibfield  {author} {\bibinfo {author} {\bibfnamefont {I.}~\bibnamefont
  {Zuriguel}}, \bibinfo {author} {\bibfnamefont {A.}~\bibnamefont
  {Garcimart{\'\i}n}}, \bibinfo {author} {\bibfnamefont {D.}~\bibnamefont
  {Maza}}, \bibinfo {author} {\bibfnamefont {L.~A.}\ \bibnamefont {Pugnaloni}},
  \ and\ \bibinfo {author} {\bibfnamefont {J.~M.}\ \bibnamefont {Pastor}},\
  }\bibfield  {title} {\enquote {\bibinfo {title} {Jamming during the discharge
  of granular matter from a silo},}\ }\href@noop {} {\bibfield  {journal}
  {\bibinfo  {journal} {Phys. Rev. E}\ }\textbf {\bibinfo {volume} {71}},\
  \bibinfo {pages} {051303} (\bibinfo {year} {2005})}\BibitemShut {NoStop}%
\bibitem [{\citenamefont {Campbell}(1989)}]{campbell1989}%
  \BibitemOpen
  \bibfield  {author} {\bibinfo {author} {\bibfnamefont {C.~S.}\ \bibnamefont
  {Campbell}},\ }\bibfield  {title} {\enquote {\bibinfo {title} {The stress
  tensor for simple shear flows of a granular material},}\ }\href@noop {}
  {\bibfield  {journal} {\bibinfo  {journal} {J. Fluid Mech.}\ }\textbf
  {\bibinfo {volume} {203}},\ \bibinfo {pages} {449--473} (\bibinfo {year}
  {1989})}\BibitemShut {NoStop}%
\bibitem [{\citenamefont {Tripathi}\ and\ \citenamefont
  {Khakhar}(2010)}]{tripathi2010}%
  \BibitemOpen
  \bibfield  {author} {\bibinfo {author} {\bibfnamefont {A.}~\bibnamefont
  {Tripathi}}\ and\ \bibinfo {author} {\bibfnamefont {D.~V.}\ \bibnamefont
  {Khakhar}},\ }\bibfield  {title} {\enquote {\bibinfo {title} {Steady flow of
  smooth, inelastic particles on a bumpy inclined plane: {H}ard and soft
  particle simulations},}\ }\href@noop {} {\bibfield  {journal} {\bibinfo
  {journal} {Phys. Rev. E}\ }\textbf {\bibinfo {volume} {81}},\ \bibinfo
  {pages} {041307} (\bibinfo {year} {2010})}\BibitemShut {NoStop}%
\end{thebibliography}%
\end{document}